\documentclass[twocolumn, tighten]{aastex63}
\usepackage{bm}
\usepackage{amsmath} 
\usepackage{booktabs}
\usepackage{array}
\usepackage{enumerate}
\usepackage{ulem,xpatch}

\xpatchcmd{\sout}
  {\bgroup}
  {\bgroup}
  {}{}

\newcommand{\KIAA}{\affiliation{Kavli Institute for Astronomy and
Astrophysics, Peking University, Beijing 100871, China}}
\newcommand{\DOA}{\affiliation{Department of Astronomy, School of Physics,
Peking University, Beijing 100871, China}}

\newcommand{\NAOC}{\affiliation{National Astronomical Observatories,
Chinese Academy of Sciences, Beijing 100012, China}}

\received{June 1, 2021}
\revised{June 1, 2021}
\accepted{June 1, 2021}
\submitjournal{AJ}
\shorttitle{LISA/TAIJI Exoplanets}
\shortauthors{Y. Kang, C. Liu, L. Shao}

\begin{document}

\title{Prospects for detecting exoplanets around double white dwarfs with LISA
and Taiji}
\correspondingauthor{Lijing Shao}
\email{lshao@pku.edu.cn}
\author[0000-0001-7402-4927]{Yacheng Kang}\DOA\KIAA
\author[0000-0001-7649-6792]{Chang Liu}\DOA\KIAA
\author[0000-0002-1334-8853]{Lijing Shao}\KIAA\NAOC


\begin{abstract}

Recently, \citet{Tamanini:2018cqb} discussed the possibility to detect
circumbinary exoplanets (CBPs) orbiting double white dwarfs (DWDs) with the
Laser Interferometer Space Antenna (LISA). Extending their methods and criteria,
we discuss the prospects for detecting exoplanets around DWDs not only by LISA,
but also by Taiji, a Chinese space-borne gravitational-wave (GW) mission which
has a slightly better sensitivity at low frequencies.  We first explore how
different binary masses and mass ratios affect the  abilities of LISA and Taiji
to detect CBPs. Second, for certain known detached DWDs with high
signal-to-noise ratios, we quantify the possibility of CBP detections around
them. Third, based on the DWD population obtained from the Mock LISA Data
Challenge, we present basic assessments of the CBP detections in our Galaxy
during a 4-year mission time for LISA and Taiji. We discuss the constraints on the detectable
zone of each system, as well as the distributions of the inner/outer edge of the
detectable zone. {Based on the DWD population, we further inject two different
planet distributions with an occurrence rate of 50\% and constrain the total
detection rates.} We finally briefly discuss the prospects for detecting
habitable CBPs around DWDs with a simplified model. These results can provide
helpful inputs for upcoming exoplanetary projects and help analyze planetary
systems after the common envelope phase.

\end{abstract}




\section{ Introduction } 
\label{ sec:intro }

\begin{figure*}[t]
    \centering
    \includegraphics[width=0.75\linewidth]{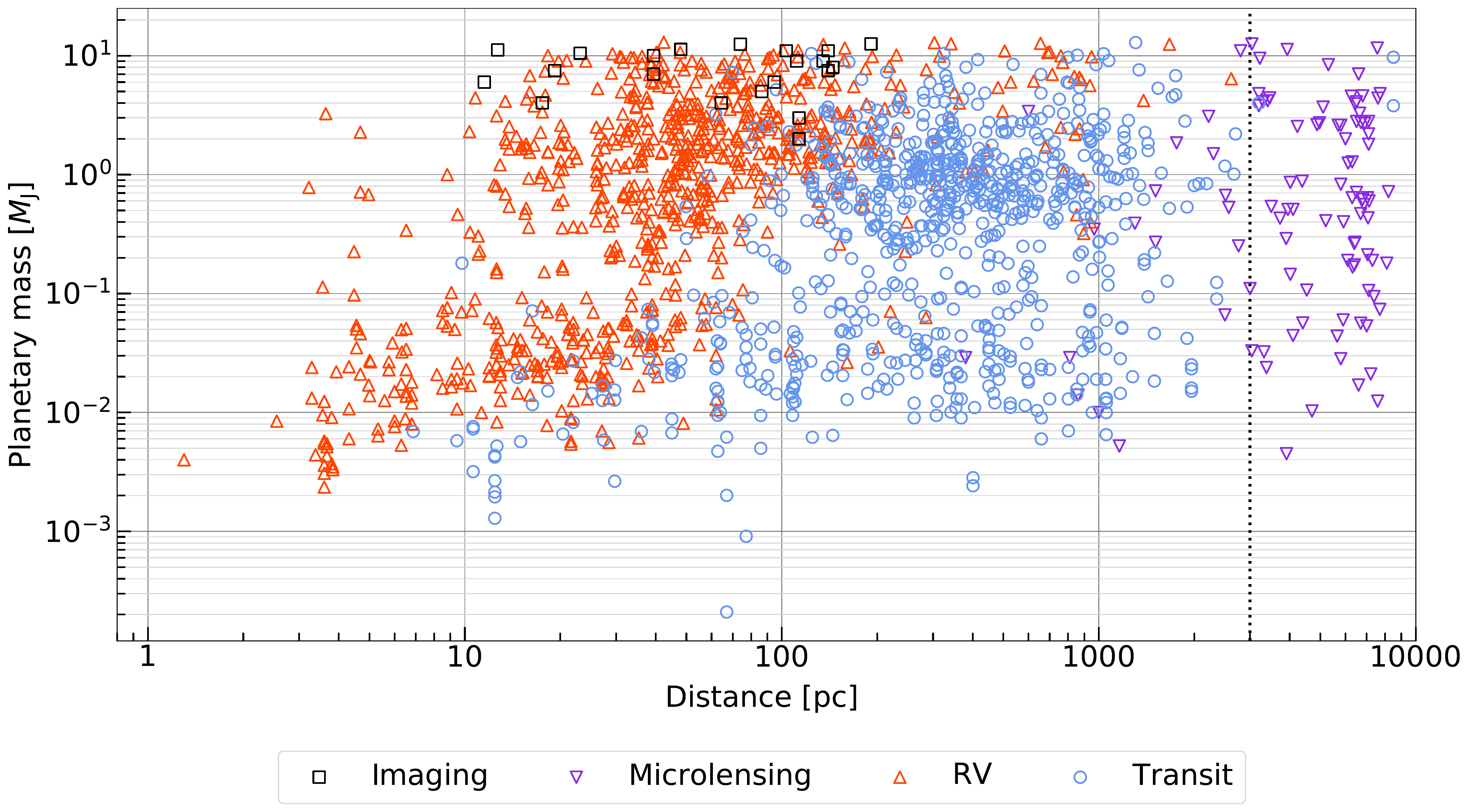}
    \caption{Distribution of planetary mass and distance of the confirmed
    exoplanets. The black dotted line marks a distance of 3\,kpc away from our
    Solar system. Different markers correspond to the currently known exoplanets
    using different EM detection techniques. Note that we only plot exoplanets
    with masses below the deuterium burning limit, i.e. $M_{\mathrm{p}} =
    13\,M_{\mathrm{J}}$ (see Sec.~\ref{ sec:Exoplanet injection }). The data
    were obtained from the NASA Exoplanet Archive. 
    \label{ fig:Exoplanets } }
\end{figure*}

So far, more than 4,300 exoplanets have been discovered using electromagnetic
(EM) techniques, but we know very little about planetary systems under extreme
conditions, such as around white dwarfs (WDs). Theoretical works suggest that a
planet can survive the host-star evolution \citep{1984MNRAS.208..763L,
1998Icar..134..303D, Nelemans:1998axa}, and the observational results also
confirm that P-type exoplanets \citep{1986A&A...167..379D} can exist around
stars after one or two common envelope (CE) phases, for example, around system
NN Ser, which contains a WD and a low-mass star \citep{Beuermann:2010bt,
Beuermann:2010ny}, and PSR B1620$-$26AB, which contains a WD and a millisecond
pulsar \citep{1993ApJ...415L..43S, 1993ApJ...412L..33T}. Nevertheless, no
exoplanets have been discovered orbiting double WDs (DWDs) to date
\citep{Tamanini:2018cqb}. Given that more than $\sim 97\%$ of stars will become
WDs \citep{Althaus:2010pi} and about 50\% of Solar-type stars are not single
\citep{Raghavan:2010hq, Duchene:2013cba}, there should be a considerable
population of DWDs in our Galaxy. If exoplanets do exist around DWDs, the
detection of such a population in the future would be very promising.  

However, even if exoplanets can endure the CE phase(s), they may collide with
each other or be ejected from evolving systems due to the complex orbital
evolution \citep{Debes:2002bx, Veras:2011di, Veras:2012yi, 2016RSOS....350571V,
2018MNRAS.476.3939M}. Strong tidal forces can crush the planetary cores during
their migration or scattering processes \citep{2018MNRAS.481.2601F}, which may
be associated with the WD pollution effect \citep{Jura:2008qm,
2016NewAR..71....9F, 2017MNRAS.468.1575B, 2018MNRAS.480...57S}. Therefore, as
noted by \citet{Danielski:2019rvt}, the detection and study of these objects can
help analyze planetary systems after CE phases and the planetary formation
processes.

Owing to the intrinsic faintness of DWDs and the sensitivity limits of the
current EM methods, there are no more than 200 known detached DWD systems
\citep{2020ApJ...889...49B}. The amount of known interacting (AM CVn) DWD
systems is even fewer \citep{2018A&A...620A.141R}. A more frustrating fact is
that most detected exoplanets are restricted to the Solar neighborhood ($\sim
3$\,kpc)  and discovered successfully by EM detection methods (see Fig.~\ref{
fig:Exoplanets }), such as radial velocity (RV) and transit
measurements.\footnote{https://exoplanetarchive.ipac.caltech.edu} Gravitational
microlensing is capable of detecting exoplanets farther away ($\sim 8$\,kpc)
towards the Galactic bulge, but scarcity and unrepeatability can be two of the
main restricting factors. All these show that it is hard to discover exoplanets
orbiting DWDs using traditional EM techniques in the Milky Way (MW).

Differently, gravitational waves (GWs) can provide a powerful tool in the
detection of exoplanets beyond our Solar system without the above selection
problem \citep{Seto:2008di, Wong:2018amf}. Recent studies have explored the prospects for
detecting new circumbinary exoplanets (CBPs) around DWDs in our Galaxy by using
the Laser Interferometer Space Antenna (LISA) mission \citep{Tamanini:2018cqb,
Danielski:2019rvt}. The method, measuring the perturbation on the GW signals due
to CBPs, is conceptually similar to the RV technique. Compared to the
traditional EM methods, GW detections are able to detect such a CBP population
in principle everywhere in the MW without being affected by stellar activities,
which, in contrast, should be considered rather carefully in EM observations. An
even more exciting prospect is that space-borne GW detectors have the potential
to detect DWDs in nearby galaxies \citep{Korol:2020lpq, Roebber:2020hso}, up to
the border of the Local Group \citep{Korol:2018ulo}. From these we can see that
in the near future, considering the rapid development of GW astronomy, the first
ever extra-galactic planetary system might be detected by the space-borne GW
detectors \citep{Danielski:2020hxb}.

In this paper, firstly, we followed the method and procedure presented in
\citet{Tamanini:2018cqb} to discuss the prospects for detecting CBPs around DWDs
by using two different space-borne GW detectors, LISA and Taiji. We give a
complementary discussion on the possibility of CBP detections around some known
detached DWDs with high signal-to-noise ratios (SNRs). Secondly, based on the
DWD population from the Mock LISA Data Challenge (MLDC) Round 4
\citep{Babak:2009cj}, we explore the population of CBP detections in our Galaxy
during a 4-year mission time for GW detectors.  For comparison, recent work has
used dedicated binary synthesis simulations for the DWD population
\citep{Korol:2017qcx, Lamberts:2019nyk}, and our results are remarkably
consistent with them, yet providing a faster way for assessments. Thirdly, we
introduce detectable zone for each promising detectable system and discuss the
distributions of the inner/outer edge of this area. {Fourthly, we inject two different planet distributions with an occurrence rate of 50\% for DWDs to constrain the
total detection rate during a 4-year mission time.} Finally, we briefly discuss
the prospects for detecting habitable CBPs around DWDs with a simplified model
by assuming that the habitable zone boundary criteria for main-sequence (MS)
stars also apply to DWDs.  These results can provide a crude benchmark for
upcoming exoplanetary projects and help analyze planetary systems after CE
phases. 

The organization of this paper is as follows. We briefly introduce the two
space-borne GW detectors that we use and the construction of their sensitivity
curves in Sec.~\ref{ sec:Detectors }. In Sec.~\ref{ sec:md }, we overview the
method proposed in \citet{Tamanini:2018cqb}, and present the characteristics of
DWD populations and CBP models used in our work for LISA and Taiji.  Using the
above ingredients, we report detailed analyses and our results on various
aspects of CBP detections in Sec.~\ref{ sec:rts }. Finally, we present a
conclusion  in Sec.~\ref{ sec:Conclusion }.


\section{ Detectors } 
\label{ sec:Detectors }

The era of GW astronomy has begun since the first direct detection of GWs,
namely GW150914 \citep{Abbott:2016blz}. Generally, the ground-based GW detectors
are sensitive to frequencies between $\sim10$\,Hz and a few of kHz, which has
made them succeed in ``listening'' to numerous GWs from merging stellar-mass
sources, like binary black holes and binary neutron stars
\citep{TheLIGOScientific:2017qsa}.  For the space-borne detectors, such as LISA
and Taiji, the sensitive frequency band ranges from 0.1\,mHz to 1\,Hz due to
their much longer arm lengths and specific optics. In such a frequency range,
the potential GW signals  come from different sources, and are considered to
have great astronomical and cosmological significances
\citep{2002grg..conf...72C, Berti:2004bd, 2016PhRvD..93b4003K,
2019PhRvD.100d4036S}. In particular, Galactic binaries including DWDs are one
class of the prominent sources emitting continuous GWs in this frequency band. 

\subsection{LISA and Taiji }
\label{ sec:2.1 }

LISA mission, proposed by an international collaboration of
scientists called the LISA Consortium, is an ESA-led L3 mission, with NASA as a
junior partner, to record and study gravitational radiation in the millihertz
frequency band \citep{Audley:2017drz}. It consists of three spacecrafts with
$2.5 \times 10^6$\,km arm-lengths trailing the Earth and moving in the Earth
orbit around the Sun. In the sensitive frequency range of LISA, the dominant GW
sources by numbers will be Galactic DWDs in the MW \citep{Lamberts:2019nyk}. So
for any CBPs orbiting DWDs, LISA would be a promising tool to indirectly detect
them. As mentioned in Sec.~\ref{ sec:intro }, the potential of LISA to detect
the first extra-galactic planetary was discussed by \citet{Danielski:2020hxb}.

On the other hand, Taiji, whose prototype was started in 2008, is a Chinese
space-borne GW mission similar to LISA. It also consists of three satellites
forming a giant equilateral triangle, but with $3 \times 10^6$\,km arm-lengths,
slightly longer than that of LISA \citep{2020ResPh..1602918L, Ruan:2020smc,
Wang:2021mou}.  These satellites are planned to orbit the Sun in the Earth orbit
with approximately 20 degrees ahead of the Earth. Taiji also aims to detect
low-frequency GW sources in the frequency band between 0.1\,mHz to 1\,Hz.  Its
sensitivity curve at the lower frequency range performs slightly better than
LISA, thus, as we will see, it has advantages on the detection of DWDs and CBPs.

\subsection{ Sensitivity curves }
\label{ sec:2.2 }

When it comes to GW detectors, sensitivity curves are important performance
guidelines. We can use them as a tool to evaluate what types of sources can be
detected during the mission. As described in \citet{Robson_2019}, we know that
the sensitivity of GW detectors depends on the GW frequency $f$, as given by
\begin{equation}
\begin{split}
S_{n}(f)=&\frac{10}{3 L^{2}}\left[P_{\mathrm{OMS}}+2\left(1+\cos ^{2}\left(f / f_{*}\right)\right) \frac{P_{\mathrm{acc}}}{(2 \pi f)^{4}}\right]\\
         &\times\left[1+\frac{6}{10}\big(\frac{f}{f_{*}}\big)^{2}\right] + S_{c}(f)\,,\label{ Snf }
\end{split}
\end{equation}
where $S_{n}(f)$ is referred to as the effective noise power spectral density;
$L$ is the arm length of space-borne GW detector; $f_{*}= c/(2 \pi L)$ is called
the transfer frequency. Due to the longer arm length of Taiji ($L = 3 \times
10^6$\,km), $f_{*}$ is a little smaller for Taiji than for LISA ($L = 2.5 \times
10^6$\,km). In the expression above, the single-link optical metrology noise is
quoted as $P_{\mathrm{acc}}$, while the single test-mass acceleration noise
$P_{\mathrm{OMS}}$ is slightly different between the two missions. More details
and analyses of these parameters can be found in \citet{Robson_2019} and
\citet{Wang:2021mou}.

Besides the instrument noise, estimates for the confusion noise $S_{c}(f)$ are
also very important. $S_{c}(f)$ is caused by the unresolved Galactic binaries,
and it is associated with the design of the space-borne detectors. As described
in \citet{Robson_2019}, estimates for $S_{c}(f)$   are well fit by
\begin{equation}
\begin{split}
S_{c}(f)=&A_{\mathrm{fix}} f^{-7 / 3} \mathrm{e}^{-f^{\alpha}+\beta f \sin (\kappa f)}\\
&\times \big[1+\tanh \big(\gamma(f_{k}-f)\big)\big] \mathrm{Hz}^{-1}\,,
\end{split}
\end{equation}
where the fixed amplitude $A_{\mathrm{fix}} = 9 \times 10^{-45}$, the knee
frequency $f_{k} = 0.00113$ and other fit parameters  are given for a 4-year
observational time as $\alpha = 0.138$, $\beta = -221$, $\kappa = 521$, and
$\gamma = 1680$.  We plot the sensitivity curves of LISA and Taiji in terms of
their characteristic strain, $\sqrt{fS_{n}(f)}$, in Fig.~\ref{ fig:SNR }. Note
that we have assumed a 4-year nominal mission duration for all the discussions
throughout this paper. The curve of Taiji taken from \citet{2020ResPh..1602918L}
is combined with $S_{c}(f)$, which we assume to be approximately the same for
LISA and Taiji due to their similar designs and configurations. In reality, as
the noise in the low-frequency band for Taiji is slightly better for LISA, such
an assumption puts us being conservative for Taiji's performance, as Taiji will
be able to distinguish more Galactic binaries at these frequencies. Future
studies could refine this point.

\section{ method } 
\label{ sec:md }

As mentioned in the Introduction, the GW method for detections of CBPs was
firstly proposed by \citet{Tamanini:2018cqb}. This approach relies on the large
DWD population with orbital periods $\lesssim 1$\,hour, which are expected to be
the most numerous GW sources for space-based mHz GW detectors
\citep{Nelemans:2001hp, Yu:2010fq, Audley:2017drz, Lamberts:2018cge,
Breivik:2019lmt}. Because of the richness of potential sources, GWs could be a
powerful tool to detect CBPs around DWDs.  In this section, we will present our
methodology to analyze the problem, which extends the original one in
\citet{Tamanini:2018cqb}.  In Sec.~\ref{ sec:DWD population }, we describe how
we obtain the DWD population with some reasonable assumptions. In Sec.~\ref{
sec:Exoplanet injection }, we provide some details about the CBP injection
process. The method for the GW detection of CBPs is discussed in Sec.~\ref{
sec:CBP detection }.

\subsection{ DWD population }
\label{ sec:DWD population }

We consider systems that compose of an exoplanet around a DWD.  For such
three-body systems, there is no doubt that the gravity of DWDs dominate the GW
signal when compared with that of exoplanets. Therefore, we provide quantitative
estimates and constraints for the detection of CBPs in our Galaxy based mainly
on the population of DWDs. To give quick assessments, we obtain the DWD
population from the MLDC Round~4, which is designed to demonstrate and encourage
the analysis of different GW sources \citep{Babak:2009cj}. MLDC Round~4 includes
a Galactic DWD population with  $\sim 3.4 \times 10^7$ interacting binaries and
$\sim 2.6 \times 10^7$ detached ones. We abandon the population of the
interacting systems mainly for two reasons: (i) the chirp masses of the
accreting systems are hard to obtain with GW observations only, and (ii)
accreting effects would complicate our analysis of CBPs. Such a treatment was
adopted in previous work as well \citep{Tamanini:2018cqb, Danielski:2019rvt},
and we leave the accreting systems for future studies.

\begin{figure}[t]
    \centering
    \includegraphics[width=0.9\linewidth]{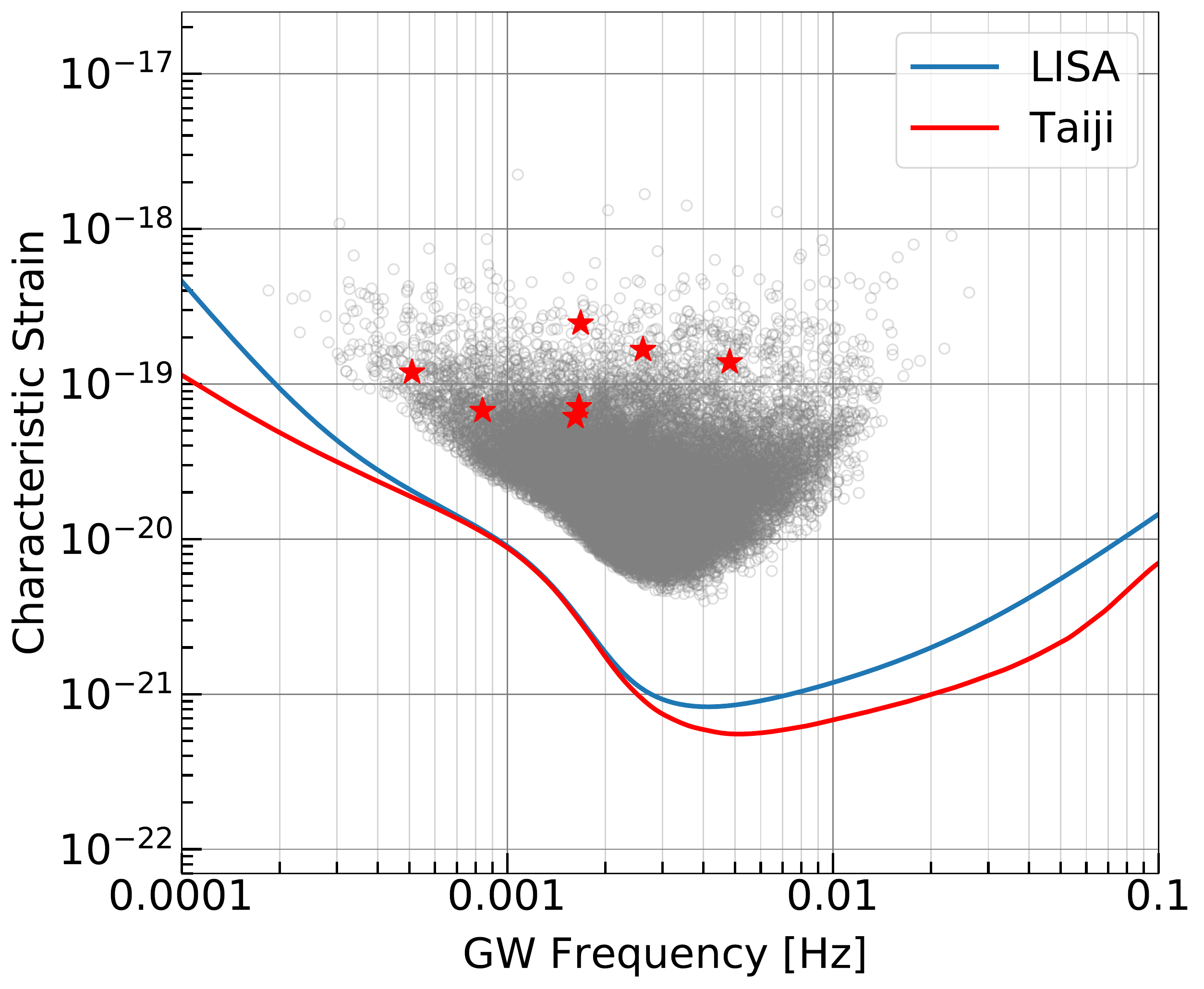}
    \caption{Characteristic strains of the detached DWD population, whose SNR $>
    10$ for Taiji, are plotted with grey circles. The known detached DWDs
    with high SNRs (listed in Table~\ref{ tab:known DWDs } and discussed in
    Sec.~\ref{ sec:4.2.1 }) are highlighted with red stars. Note that the DWD
    population is plotted based on a crude assumption that two WDs are equal in
    mass. We also plot the sensitivity curves of LISA (blue line) and Taiji (red
    line) as given in Sec.~\ref{ sec:2.2 }.}
    \label{ fig:SNR }
\end{figure}

Different values of DWD parameters not only lead to different SNRs, but also
change the estimations of detecting abilities. This was analyzed in
\citet{Tamanini:2018cqb}. We will also discuss the detecting abilities with
different values of mass and mass ratio in Sec.~\ref{ sec:4.1 }. When we perform
the calculations to assess the prospects of the final CBP detections in the MW,
we acquire the parameters of each binary from the
dataset,\footnote{\url{https://asd.gsfc.nasa.gov/archive/astrogravs/docs/mldc/}}
including the GW frequency $f$, the frequency derivative $\dot{f}$, the ecliptic
latitude $\beta$, the ecliptic longitude $\lambda$, the GW amplitude
$\mathcal{A}$, the inclination $\iota$, and the polarization phase $\psi$. More
details of the population and the parameters are presented in
\citet{Babak:2008aa, Babak:2009cj}. As we will see later in Sec.~\ref{ sec:3.3.1
}, we can derive the chirp mass $\mathcal{M}$ of each system through the use of
observed GW frequency $f$ and its time derivative $\dot{f}$. By assuming that
the two WDs are almost equal in mass, we can then acquire the mass of each WD
pair, ($m_{1}$,$m_{2}$), and their total mass $M_\mathrm{b} \equiv m_1 + m_2$.
We regard this as a crude but reasonable treatment because the mass ratio
$q=m_{1} / m_{2}$ ($m_{1} > m_{2}$) under discussion is often considered lower
than 3 for detected DWDs \citep[for example, see e.g.][]{Korol:2017qcx}. In
Sec.~\ref{ sec:4.1 }, we will show little differences in detecting abilities
when realistic mass ratio  is considered. 

With the above consideration, we re-calculate the SNRs of 31,530 ``bright"
detached Galactic binaries from MLDC Round 4 for the two detectors. We find that
approximately $2.9 \times 10^4$ ($2.2 \times 10^4$) detached DWDs have
$\mathrm{SNR} > 7$ for Taiji (LISA) during a 4-year mission. The number becomes
$2.5 \times 10^4$ ($1.6 \times 10^4$) for $\mathrm{SNR} > 10$. For the
following, we filter out all detached binaries with $\mathrm{SNR} < 10$,  for
both LISA and Taiji, to get more reliable estimations and striking contrasts
between the two missions in detecting abilities.  Also, a high SNR is generally
needed in order to have CBP detections around DWDs.  In Fig.~\ref{ fig:SNR } we
show the dimensionless characteristic strain of the detached DWD population with
$\mathrm{SNR} > 10$ for Taiji. Because of the direct use of the DWD catalog, we
could provide faster assessments for comparisons which are consistent with
previous work within an order of magnitude \citep{Korol:2017qcx,
Lamberts:2019nyk, Danielski:2019rvt}.

\subsection{ Injection of CBP models }
\label{ sec:Exoplanet injection }

When we consider CBPs around a DWD, there is no evidence to claim that every DWD
should have such an exoplanet. Given that no planets have been discovered
orbiting DWDs so far, we take a bold approach, following
\citet{Danielski:2019rvt}, and set 50\% as the occurrence rate for our synthetic
population of CBPs around DWDs, which is obtained according to the observed
frequency of WD pollution effect \citep{2014A&A...566A..34K}. Note also that
even if such CBPs exist, we may miss these exoplanets using space-borne GW
detectors for a variety of reasons. {Therefore, a combination of diverse
semi-major axis ($a$) and CBP mass ($M_\mathrm{p}$) distributions have been
tested in \citet{Danielski:2019rvt}, from which we adopt the optimistic and the pessimistic cases as reference points. Notice that there is a difference between the distributions in \citet{Danielski:2019rvt} and ours in the planet's orbital inclination $i$. Instead of setting a uniform distribution in $\cos i$, we inject CBPs into the DWD systems by assuming coplanar circular orbits. There are theoretical indications that CBPs prefer to be coplanar with their central binaries \citep{Kennedy:2012gc, Foucart:2012xe}, and coplanar orbits have been considered in various other work as well \citep{1986A&A...167..379D, 1999AJ....117..621H, 2008A&A...489.1329E,2021NewA...8401516H}. It is certainly advantageous to refine the currently quite uncertain CBP population models in future for more accurate predictions, in particular for a realistic estimate via the GW method. We will show more details and our detection rates in Sec.~\ref{ sec:different distributions }. } 

\subsection{ Detection of CBPs around DWDs }
\label{ sec:CBP detection }

This subsection briefly introduces the method for the GW detection of a CBP
using space-borne GW detectors. We follow \citet{Tamanini:2018cqb} to model the
perturbation induced by CBPs around DWDs.  We first describe some
characteristics of the three-body system in Sec.~\ref{ sec:3.3.1 }, and then
provide  more details about the parameter estimation process in Sec.~\ref{
sec:3.3.2 }.

\subsubsection{ Perturbation due to a CBP }
\label{ sec:3.3.1 }

Considering a three-body system composed of a DWD emitting GWs with an exoplanet
on the outer orbit (P-type system), we assume that the separation between the
planet and the DWD is much greater than the separation between the two WDs. For
simplicity, we also consider both these orbits as circular Keplerian orbits. 
This could root in the binary evolution scenarios.  Based on these assumptions,
we obtain the radial velocity of the DWD with respect to the common center of
mass (CoM), 
\begin{equation}
v_{\mathrm{rad}}(t)=-K \cos \varphi(t)\,.
\label{eq:v}
\end{equation}
We have defined,
\begin{align}
K &=\Big(\frac{2 \pi G}{P}\Big)^{1/3} \frac{M_{\mathrm{p}}}{\left(M_{\mathrm{b}}+M_{\mathrm{p}}\right)^{2/3}} \sin i\,,\label{ eq:K } \\
\quad \varphi(t) &=\frac{2 \pi t}{P}+\varphi_{0}\,,
\end{align}
where $P$ and $i$ are respectively the orbital period and inclination of the
CBP, $\varphi(t)$ is the outer orbital phase, and $\varphi_0$ is its initial
value at $t=0$.

Through the Doppler effect, the observed GW frequency changes in the Earth
reference frame to,
\begin{equation}
f_\mathrm{o b s}(t)=\left(1+\frac{v_\mathrm{rad}(t)}{c}\right) f_\mathrm{G W}(t)\,,
\end{equation}
where $f_\mathrm{G W}(t)$ is the GW frequency in the DWD reference frame (twice
the DWD orbital frequency). Galactic binaries take  much longer than the mission time of GW detectors to
merge, and  their frequencies are changing very slowly.  We can then describe
their time evolution with a Taylor expansion and neglect the second and
higher-order terms by using,
\begin{equation}
f_\mathrm{G W}(t)=f+\dot{f} t+O\left(t^{2}\right)\,,
\end{equation}
where $f$ is the initial observed GW frequency, and $\dot{f}$ is its time
derivative, which are related to the chirp mass $\mathcal{M}=(m_{1} m_{2})^{3 /
5}/(m_{1}+m_{2})^{1/5}$ of the DWD system via,
\begin{equation}
\dot{f}=\frac{96}{5} \pi^{8 / 3} f^{\ 11 / 3}\left(\frac{G \mathcal{M}}{c^{3}}\right)^{5 / 3}\,.
\label{ eq:f_dot }
\end{equation}
Finally, by integrating the observed GW frequency $f_\mathrm{obs}(t)$, we can
obtain the phase at the observer of the GWs,
\begin{equation}
\Psi_\mathrm{o b s}(t)=2 \pi \int f_\mathrm{o b s}\left(t^{\prime}\right) d t^{\prime}+\Psi_{0}\,,
\end{equation}
where $\Psi_{0}$ is the constant initial phase.  The final form of the observed
phase is given by,
\begin{equation}
\begin{split}
\Psi_\mathrm{o b s}(t) =& 2 \pi\Big(f+\frac{1}{2} \dot{f} t\Big) t-  \frac{
P f}{c} K \sin \varphi(t)\\
                & - \frac{ P \dot{f} t }{c}  K \sin \varphi(t)- \frac{ P^{2}
                \dot{f} }{2 \pi c} K \cos \varphi(t)\,.
\end{split}
\end{equation}
With all the equations above, the parameters characterizing the DWD and the
perturbation induced by a CBP can thus be extracted from the GW phase evolution.

\subsubsection{ Parameter estimation for LISA and Taiji }
\label{ sec:3.3.2 }

In low frequency range, LISA and Taiji each can be effectively seen as a pair of
two-arm GW detectors like LIGO and Virgo, and output two linearly independent
signals, $h_{\mathrm{I}}(t)$ and $h_{\mathrm{II}}(t)$. We often assume that the
noise is stationary and Gaussian, and then the two signals in each independent
channel can be written as \citep{1998PhRvD..57.7089C},
\begin{equation}\label{eq:ht}
h_\mathrm{I, I I}(t) =\frac{\sqrt{3}}{2} A_\mathrm{I, I I}(t) \cos
\big[\Psi_\mathrm{o b s}(t) +\Phi_\mathrm{I, I
I}^{\mathrm{p}}(t)+\Phi_\mathrm{D}(t)\big]\,,
\end{equation}
where $A_\mathrm{I, I I}(t)$ are amplitudes of GW signals that contain the
constant intrinsic amplitudes of the waveform and the antenna pattern functions
of the detector. In our case the waveform is approximated by a circular
Newtonian binary. The antenna pattern functions are related to geometric
parameters, including the location of the source ($\theta_{S}, \phi_{S}$), the
orientation of the DWD orbit ($\theta_{L}, \phi_{L}$), and the configuration of
the space-borne detector.  In Eq.~(\ref{eq:ht}), $\Phi_\mathrm{I, I
I}^{\mathrm{p}}(t)$ are the waveform's polarization phases induced by the change
of the orientation of the detector.  The Doppler phase $\Phi_\mathrm{D}(t)$ is
the difference between the phase of the wavefront at the detector and the phase
of the wavefront at the Sun. It is further related to the Earth-Sun distance and
the orbital period of the Earth.  The full expressions for all above quantities
can be found in \citet{1998PhRvD..57.7089C}, \citet{Cornish:2003vj}, and
\citet{Korol:2017qcx}. 

Based on the above analysis, our next step is to simulate the response of LISA
and Taiji and perform parameter estimation. We use the Fisher information
approach, as was employed by \citet{Tamanini:2018cqb}. For each DWD, there are
11 parameters, $\boldsymbol \lambda = \big\{ \ln (A), \Psi_{0}, f, \dot{f},
\theta_{S}, \phi_{S}, \theta_{L}, \phi_{L}, K, P, \varphi_{0} \big\}$,
characterizing the observed GW waveform. The Fisher matrix can be written as
\begin{equation}
\Gamma_{i j}=\frac{2}{S_{n}\left(f\right)} \sum_{\alpha=\mathrm{I, I I}}
\int_{0}^{T_\mathrm{o b s}}  \left(\frac{\partial h_{\alpha}(t)}{\partial
\lambda_{i}} \cdot \frac{\partial h_{\alpha}(t)}{\partial \lambda_{j}}\right)
d t\,.
\label{ eq:Gamma }
\end{equation}
We use the one-sided noise power spectral density $S_{n}(f)$ of the detector
from Eq.~(\ref{ Snf }). For each DWD, it is merely a constant in Eq.~(\ref{
eq:Gamma }) because the binary is quasi-monochromatic during the observational
time $T_\mathrm{obs}$ as long as $\dot{f} T_\mathrm{obs} \ll f$.

Similarly, the SNR of the signal can be written as,
\begin{equation}
\mathrm{SNR}= \bigg( \frac{2}{S_{n}\left(f\right)} \sum_{\alpha=\mathrm{I, I I}}
\int_{0}^{T_{\rm obs}} h_{\alpha}(t) h_{\alpha}(t)  \, d t \bigg)^{1/2} \,.
\end{equation}
This allows us to scale all results with the SNR by rescaling $S_{n}(f)$
\citep{Tamanini:2018cqb}. From the inverse of the Fisher matrix, we can obtain
the uncertainties and correlations of  parameters as the elements of the
variance-covariance matrix $\Sigma_{i j}$,
\begin{equation}
\Sigma_{i j}=\left\langle\Delta \lambda_{i} \Delta \lambda_{j}\right\rangle=\left(\Gamma^{-1}\right)_{i j}\label{ eq:Sigma }\,.
\end{equation}

\citet{1998PhRvD..57.7089C} has studied the uncertainties for binary parameters.
We follow his method and determine different partial derivatives of
$\Psi_\mathrm{o b s}(t)$. Since the error $\Delta \dot{f}$ would be much larger
than the signal's $\dot{f}$ itself,  we adopt a treatment to simply set the
fiducial value $\dot{f}=0$ without introducing noticeable changes
\citep{Takahashi:2002ky}. Here we only show expressions of the partial
derivatives differing from the equations in the Sec.\,$\text{IV}$ of
\citet{1998PhRvD..57.7089C},
\begin{equation}
\begin{split}
\partial_{K} \Psi_\mathrm{o b s}(t)&=-\frac{P f}{c} \sin \varphi(t)\,,
\\
\partial_{P} \Psi_\mathrm{o b s}(t)&=-\frac{ f}{c} K \sin \varphi(t)+\frac{2 \pi  ft }{cP} K \cos  \varphi(t)\,,
\\
\partial_{f}\: \Psi_\mathrm{o b s}(t)&=2 \pi t-\frac{ P}{c} K \sin \varphi(t)\,,
\\
\partial_{\dot{f}}\: \Psi_\mathrm{o b s}(t)&=\pi t^{2}-\frac{ P t}{c} K \sin \varphi(t)-\frac{P^{2}}{2 \pi c} K \cos \varphi(t)\,,
\\
\partial_{\varphi_0} \Psi_\mathrm{o b s}(t)&=-\frac{ P f}{c} K \cos \varphi(t)\,.
\end{split}
\end{equation}

To measure the additional perturbation due to an exoplanet,
\citet{Tamanini:2018cqb} paid more attention to the three parameters associated
with the CBP, namely $K, P, \varphi_{0}$. Since the value of $\varphi_{0}$ is
not important for our final results, we  fix $\varphi_{0} = \pi /2$. We set 30\%
to be the detection criterion on both $\Delta K/K$ and $\Delta P/P$, meaning
that a detectable CBP is defined with estimated parameter precision being better
than this value.

\section{ Results } 
\label{ sec:rts }

Now we give detailed analyses and discuss the prospects for detecting exoplanets
around DWDs. Using Taiji as a demonstration, we first give some complementary
discussion on the detecting abilities with different values of masses in
Sec.~\ref{ sec:4.1 }. We compare our final results of the CBP detection between
LISA and Taiji in Sec.~\ref{ sec:Comparison }. 

\subsection{ Effects of masses}
\label{ sec:4.1 }

\begin{figure}
    \centering
    \includegraphics[width=0.9\linewidth]{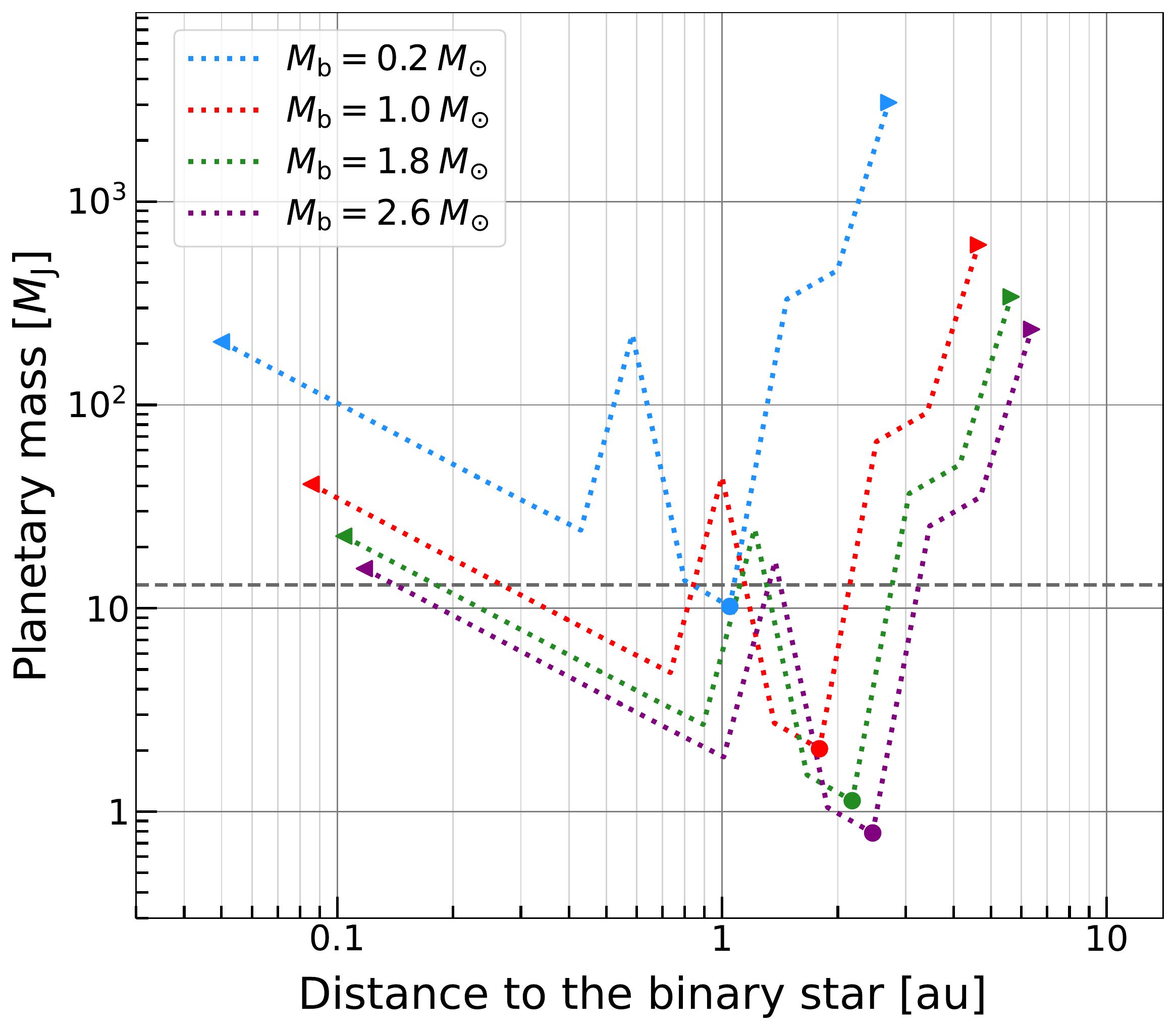}
    \caption{Selection functions of Taiji in the mass-separation parameter space
    for exoplanets. Four dotted lines in different colors denote systems with
    different total masses. The dashed horizontal grey line corresponds to the
    deuterium burning limit $M_{\mathrm{p}} = 13$\,$ M_{\mathrm{J}}$. Circles
    are the detectable minimum planetary masses for each system, and left/right
    triangles denote the boundary values we set. The peak of each three-body
    system is caused by the degeneracy between the motion of Taiji and the
    motion of the DWD around the three-body CoM.} 
    \label{ fig:Different mass }
\end{figure}

From Eq.~(\ref{ eq:K }), we can see that when other parameters related to the
CBP are fixed, such as its angular position and the distance, the perturbation
$K$ caused by the exoplanet is getting smaller with the increase of the total
mass $M_{\mathrm{b}}$. Meanwhile, the SNR will be in contrast higher when we
only increase the total mass of the DWD. Therefore, to give a quantitative
evaluation, we perform parameter estimation on the same set of systems except
that their total masses are ranging from 0.2\,${M}_{\odot}$ to
2.8\,${M}_{\odot}$. We assume two WDs are equal in mass and fix the other
parameters of the DWD as
\begin{equation}
\Psi_{0}=0 \,, \quad 
\theta_{S}=1.27 \,, \quad 
\varphi_{S}=5 \,, \quad 
i=\iota=\frac{\pi}{3}\,.
\end{equation}
The frequency and distance of the system are fixed to $f = 5$\,$\mathrm{mHz}$ and
$d_{\mathrm{DWD}}$ = 10\,kpc respectively. Note that we derive the orientation
of its orbit ($\theta_{L},\varphi_{L}$) by inversing Eq.~(39) and Eq.~(40) in
\citet{Cornish:2003vj}. 

We plot for comparison in Fig.~\ref{ fig:Different mass } the selection
functions of Taiji based on these values. {The four dotted lines in different
colors denote DWDs with different total masses, and the dashed horizontal grey line denotes the deuterium burning limit $M_{\mathrm{p}} = 13$\,$M_{\mathrm{J}}$, which is considered to be the upper limit of the CBP mass \citep{Danielski:2019rvt}.} Therefore, it is the area above the dotted lines and below the
dashed line that delimits the detectable mass-separation parameter space of the
CBP for Taiji. The peak of each line is caused by the degeneracies between the
motion of Taiji, and the motion of the DWD in the three-body system around its
common CoM \citep[see the section ``Methods'' in][]{Tamanini:2018cqb}. Sharp
peaks appear at $P = 1$\,yr (the orbital period of Taiji), while there are also
other smaller peaks at higher harmonics of it. 
 
\begin{figure}[t]
    \centering
    \includegraphics[width=\linewidth]{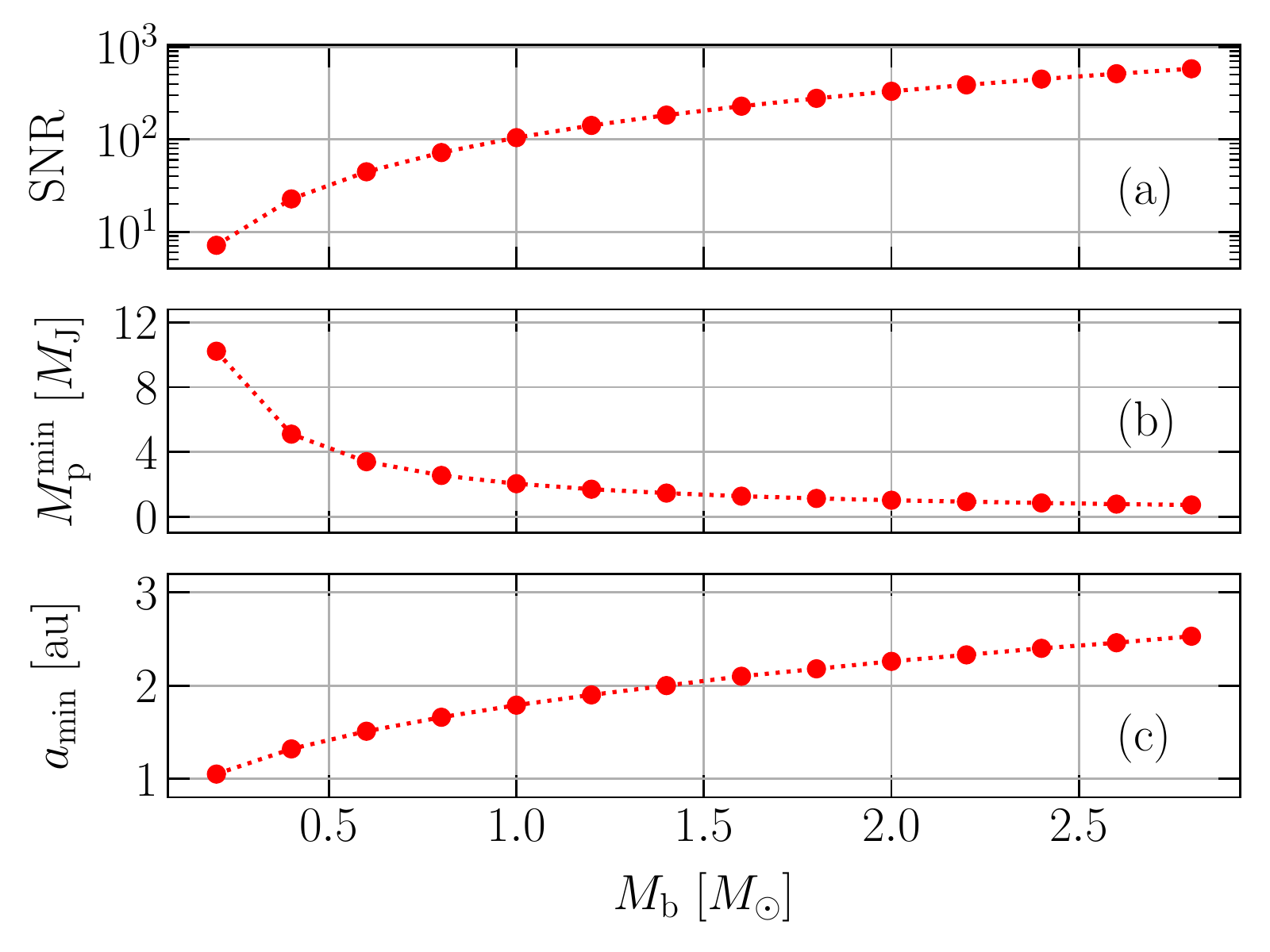}
    \includegraphics[width=\linewidth]{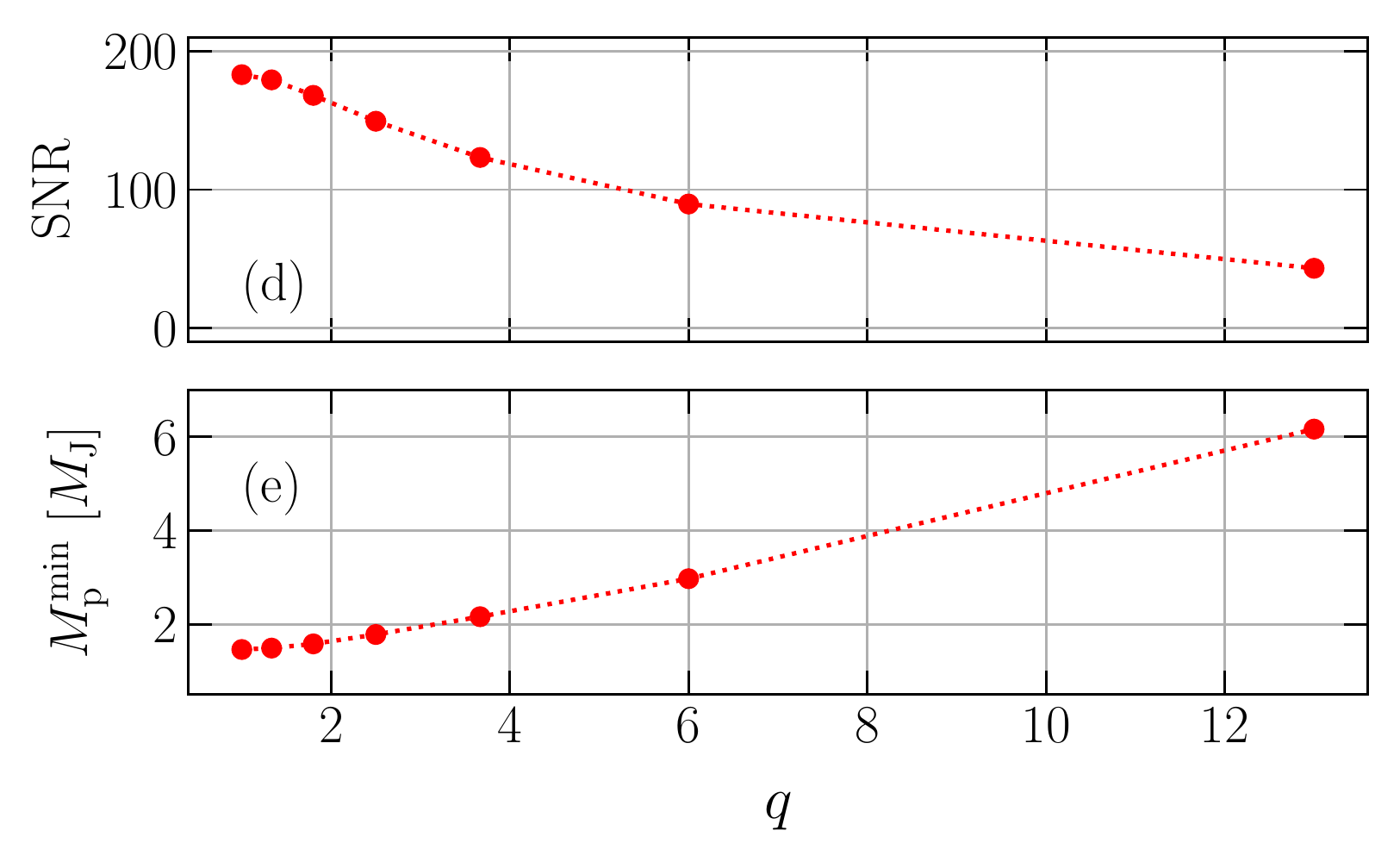}
    \caption{(a) The SNR, (b) the detectable minimum planetary mass, and (c) its
    corresponding orbital size, as functions of the total mass for equal-mass
    DWDs. (d) The SNR and (e) the detectable minimum planetary mass, as
    functions of the mass ratio $q$ for DWDs with a total mass $M_{\rm b}=
    1.4\,M_{\odot}$. More details about the parameters are given in Sec.~\ref{
    sec:4.1 }.}
    \label{fig:varying:Mb:q}
\end{figure}

For each system, we have fixed the range of the CBP orbital period to $P=
0.025$--$10$\,yr, and calculated the CBP's distance to the CoM by Kepler's
third law,
\begin{equation}
a^{3}=\frac{GP^{2}}{4 \pi^{2}}(M_\mathrm{b}+M_\mathrm{p})\,.
\label{ eq:Kepler’s law }
\end{equation}
We mark the distance boundaries as the left and right triangles in Fig.~\ref{
fig:Different mass }. When the period of the CBP is set to a same value, the
planetary orbital size gets larger with the increase of the total mass of the
DWD. We find that detecting abilities of each system are gradually getting
better with the increasing CBP period before becoming worse near the period of
one year. There must be a minimum value $M_\mathrm{p}^\mathrm{min}$ that
corresponds to the detectable minimum planetary mass for each system, which we
denote by circles in Fig.~\ref{ fig:Different mass }. 

\begin{table*}[]
    \centering
    \caption{Possible detections of CBPs around known DWDs. For each system,
    $a_\mathrm{min}$ is the detectable minimum planetary orbital size based on
    our parameter estimation. Through the comparative results of LISA and Taiji,
    we list the detectable minimum planetary mass
    $M_{\mathrm{p}}^{\mathrm{min}}$ with the SNR for each DWD. For each
    promising system, we calculate the IDZ and ODZ. Other relevant properties of
    these known detached DWDs are in the Appendix.}
    \setlength{\tabcolsep}{2.8mm}{\begin{tabular}{l r@{.}l r@{.}l cc r@{.}l r@{.}l cc r@{.}l}
    \toprule
              & \multicolumn{2}{c}{}
              & \multicolumn{5}{c}{\hspace{1.5em}Taiji}    
              & \multicolumn{5}{c}{\hspace{6em}LISA}\\
    Source    & \multicolumn{2}{c}{$a_\mathrm{min}$}
              & \multicolumn{2}{c}{\hspace{2em}$M_{\mathrm{p}}^{\mathrm{min}}$}        
              & IDZ    & ODZ    & \multicolumn{2}{c}{SNR}
              & \multicolumn{2}{c}{\hspace{1.0em}$M_{\mathrm{p}}^{\mathrm{min}}$}        
              & IDZ    & ODZ    & \multicolumn{2}{c}{SNR}          
              \vspace{-0.3em}\\
              & \multicolumn{2}{c}{[au]}
              & \multicolumn{2}{c}{\hspace{2em}[$M_\mathrm{J}$]}               
              & [au]    & [au]    & \multicolumn{2}{c}{}
              & \multicolumn{2}{c}{\hspace{1.0em}[$M_\mathrm{J}$]}               
              & [au]    & [au]    & \multicolumn{2}{c}{}                
              \vspace{0.3em}\\
    \midrule
    ZTF~J153932.16+502738.8    & 1    & 53    &\hspace{1.5em}1    & 39    
                               & 0.20    & 2.20
                               & 124  & 51    
                               & \hspace{0.5em}2        & 11        
                               & 0.32 & 2.12
                               & 81  & 89   \\
    SDSS~J065133.34+284423.4   & 1    & 79    &\hspace{1.5em}2    & 34         
                               & 0.31    & 2.04
                               & 117  & 18    
                               & \hspace{0.5em}2        & 95        
                               & 0.49 & 2.00
                               & 92   & 83    \\
    SDSS~J093506.92+441107.0   &2     & 03    &\hspace{1.5em}4    & 76        
                               & 0.73    & 2.21
                               & 139  & 66    
                               & \hspace{0.5em}9        & 50        
                               & 1.40 & 2.12
                               & 70   & 02   \\
    SDSS~J232230.20+050942.06  & 1    & 59    &\hspace{1.5em}10   & 47        
                               & 1.35    & 1.66 
                               & 70   & 72
                               & \hspace{0.5em}21       & 27        
                               & --   & --
                               & 34   & 82    \\
    PTF~J053332.05+020911.6    & 1    & 63    &\hspace{1.5em}17   & 94        
                               & --      & -- 
                               & 29   & 09
                               & \hspace{0.5em}37       & 69        
                               & --   & --
                               & 13   & 84    \\
    SDSS~J163030.58+423305.7   &1     & 72    &\hspace{1.5em}104  & 33        
                               & --      & -- 
                               & 12   & 38
                               & \hspace{0.5em}283      & 92        
                               & --   & --
                               & 4    & 55    \\
    SDSS~J092345.59+302805.0   &2     & 06    &\hspace{1.5em}179  & 01        
                               & --      & -- 
                               & 11   & 61
                               & \hspace{0.5em}412      & 89        
                               & --   & --
                               & 5    & 03    \vspace{0.5em}\\
    \bottomrule
    \end{tabular}}
    \label{ tab:known DWDs comparison }
\end{table*}

Moreover, we can see that the detectable parameter space is getting wider with
the increase of the total mass of a binary, and the detectable minimum planetary
mass is getting smaller. We point out that this can be explained by the
parameter estimation criterion on $K$, since in most cases the 30\% criterion is
not applicable for $\Delta P/P$, as we can see in the Fig.~2 of
\citet{Tamanini:2018cqb}. Eq.~(\ref{ eq:K }) tells us that the amplitude $K
\propto M_\mathrm{b}^{-2/3}$. After plugging it into Eq.~(\ref{ eq:Gamma }) and
Eq.~(\ref{ eq:Sigma }), we get $\Delta K \propto M_\mathrm{b}^{-5/3}$.
Therefore, the detectable minimum planetary masses limited by $\Delta K/K$ are
inversely proportional to the total masses of DWDs. We also find similar results
in the upper panels of Fig.~\ref{fig:varying:Mb:q} with a wider range of $M_b$
from 0.2\,$M_{\odot}$ to 2.8\,$M_{\odot}$. From these we conclude that SNR is
the dominating factor in detecting abilities when we change the total mass of
the DWD. Note that we choose 2.8\,$M_{\odot}$ to be the upper limit in
this section because we notice that the maximum total mass can reach
2.8\,$M_{\odot}$ when we considered a Galactic DWD population from MLDC Round 4
with an equal-mass assumption in the following sections. Although
2.8\,$M_{\odot}$ may be too high to be the upper limit for most DWDs, whose
total masses usually do not exceed 2\,$M_{\odot}$ \citep{Lamberts:2019nyk}, it
would not alter the qualitative conclusions derived here.

Similarly, by changing the mass ratio $q$ with a fixed total mass $M_{\rm b} =
1.4\,M_{\odot}$ of the DWD, we illustrate our results in the lower panels of
Fig.~\ref{fig:varying:Mb:q}. It supports that the detecting abilities are not
weakened too much if the deviation of $q$ from our assumption (equal mass) is
reasonable, say, $q \lesssim 3$. Based on this, we claim with confidence that our
main results have captured the major features of CBP detections.

\subsection{ Comparison between LISA and Taiji }
\label{ sec:Comparison }

Now we  compare our results about the detections of CBPs orbiting DWDs between
LISA and Taiji. We first discuss the possible detections around known DWDs in
Sec.~\ref{ sec:4.2.1 }. We find the detectable minimum planetary mass
$M_\mathrm{p}^\mathrm{min}$ of each system to see if it is below the deuterium
burning limit $M_{\mathrm{p}} = 13$\,$ M_{\mathrm{J}}$. If so, we will define
this system as the \emph{promising system}. In Sec.~\ref{ sec:4.2.2 }, we show
the distribution of the promising systems in our Galaxy and discuss the
constraint on their detectable zones (DZs), which are described as the
circumbinary distances where the space-borne GW detector has the possibility to
detect CBPs with $M_{\mathrm{p}} \le 13\,M_{\mathrm{J}}$. The distributions of
the inner/outer edge of the DZ (referred to as IDZ/ODZ) and their dependence on
the GW frequency are plotted for comparison between LISA and Taiji.
In Sec.~\ref{ sec:different distributions }, we analyze the detection rates
during 4 years by injecting different planet distributions. Finally, we discuss
the prospects for the detection of CBPs in habitable zones (HZ) around DWDs in
Sec.~\ref{ sec:HZ }.

\subsubsection{ Possible detections of CBPs around known DWDs }
\label{ sec:4.2.1 }

We first discuss possible detections of CBPs around the known detached DWDs with
high SNRs. \citet{Danielski:2019rvt} have analyzed one DWD in detail
(ZTF~J153932.16+502738.8). Here we calculate the expected SNRs for all DWDs in
\citet{Huang:2020rjf} for LISA and Taiji, and list the results of the ones with
high SNRs ($\mathrm{SNR} > 10$ for Taiji) in Table~\ref{ tab:known DWDs
comparison }. We can see that there are four promising systems for Taiji:
ZTF~J153932.16+502738.8, SDSS~J065133.34+284423.4, SDSS~J093506.92+441107.0, and
SDSS~J093506.92+441107.0, of which the first three are also promising systems
for LISA. Comparing the results in Table~\ref{ tab:known DWDs comparison },
Taiji obviously has a better performance on detecting abilities from two
aspects: (i)  the smaller detectable minimum planetary masses
$M_\mathrm{p}^\mathrm{min}$, and (ii) the wider DZs. This mainly comes from the
noticeable differences in SNRs between the two missions. 

As mentioned in Sec.~\ref{ sec:Exoplanet injection }, we consider coplanar
circular orbit for the CBP in a three-body system. Therefore, despite it seems
likely that LISA and Taiji can detect exoplanets down to $\sim
1$\,$M_{\mathrm{J}}$ around these known detached DWDs, the planetary orbital
inclinations can in fact deviate from the DWD inclination $\iota$, which may
lead to a raise in $M_\mathrm{p}^\mathrm{min}$. This could happen due to the
degeneracy between the planetary mass and inclination. On the other hand, if
complementary EM observations in the future could constrain the planetary
inclination well, we can then derive bounds for the mass of the CBP. Conversely,
space-borne GW detectors could also give constraints in the mass-separation
parameter space (see e.g., Fig.~\ref{ fig:Different mass }), and our results can
provide inputs for the EM exoplanetary projects, which are especially desirable
for the study on possible synergy between GW and EM observations.

\begin{figure*}[htbp]
    \centering
    \includegraphics[width=0.8\linewidth]{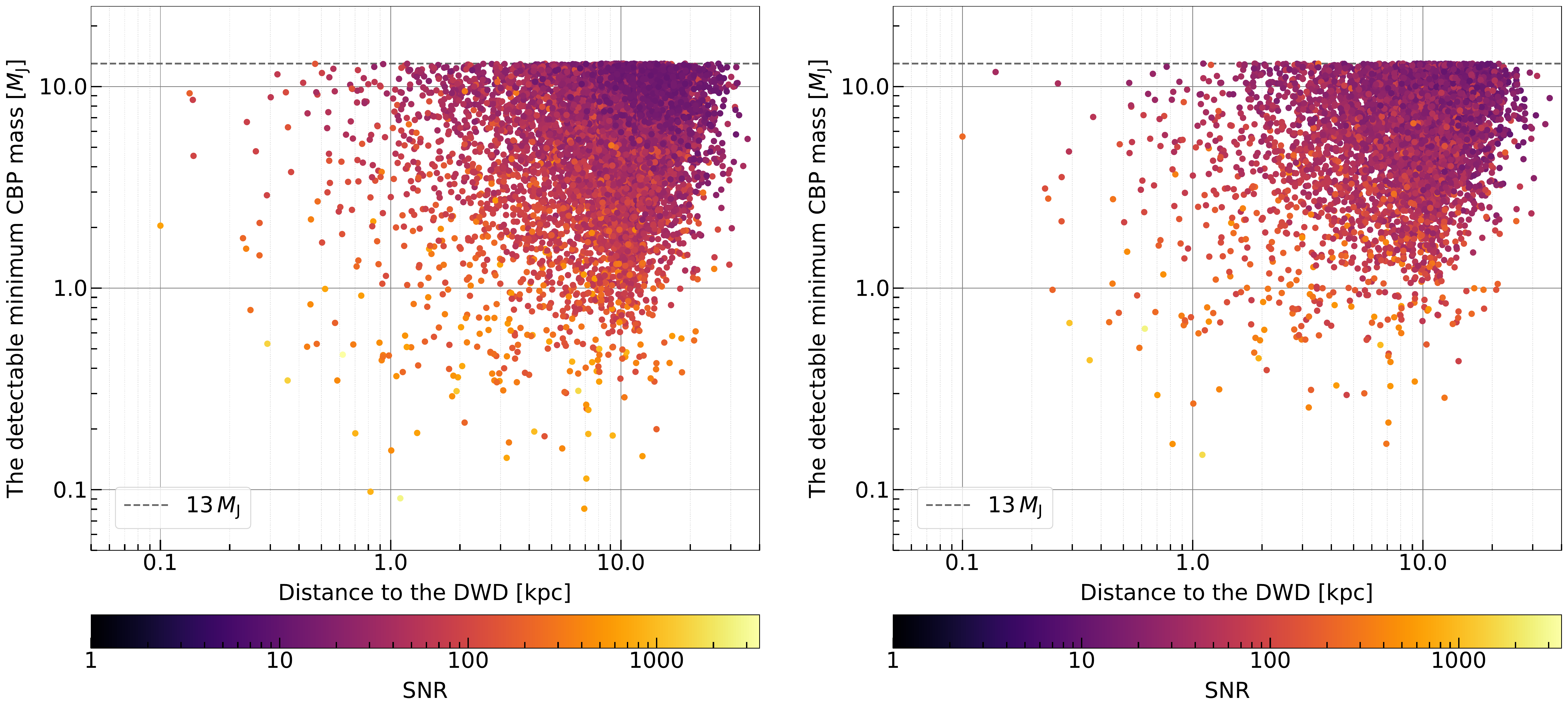}
    \caption{The relationship between the distance and the detectable minimum
    CBP mass for Taiji (left) and LISA (right)  for promising systems in MLDC
    Round 4 during a nominal 4-year mission. The color represents the SNR of
    each system, and the dashed horizontal grey line corresponds to the
    deuterium burning limit.  }
    \label{ fig:Distribution_1 }
\end{figure*}

\subsubsection{ Constraints on the promising systems in our Galaxy }
\label{ sec:4.2.2 }

We mentioned in Sec.~\ref{ sec:DWD population } that $\mathrm{SNR} > 10$ is
chosen to be the threshold for our work, and in MLDC Round~4 there are 25,016
(15,903) detached DWDs satisfying this criterion for Taiji (LISA). Based on
these populations, we find a total of 9,053 (6,718) {\it promising systems} at
most during a 4-year observation  for Taiji (LISA). Figure~\ref{
fig:Distribution_1 } shows that most systems are clustered together in the
1--13\,$M_\mathrm{J}$ mass range and at about 10\,kpc from our Solar system,
which is consistent with the distance to the Galactic center. Generally, nearby
DWDs could have lower $M_\mathrm{p}^\mathrm{min}$ than distant systems due to
their higher SNRs. This also explains why the promising systems are mainly
distributed in the top right of Fig.~\ref{ fig:Distribution_1 }. 

\begin{figure*}[htbp]
    \centering
    \includegraphics[width=0.8\linewidth]{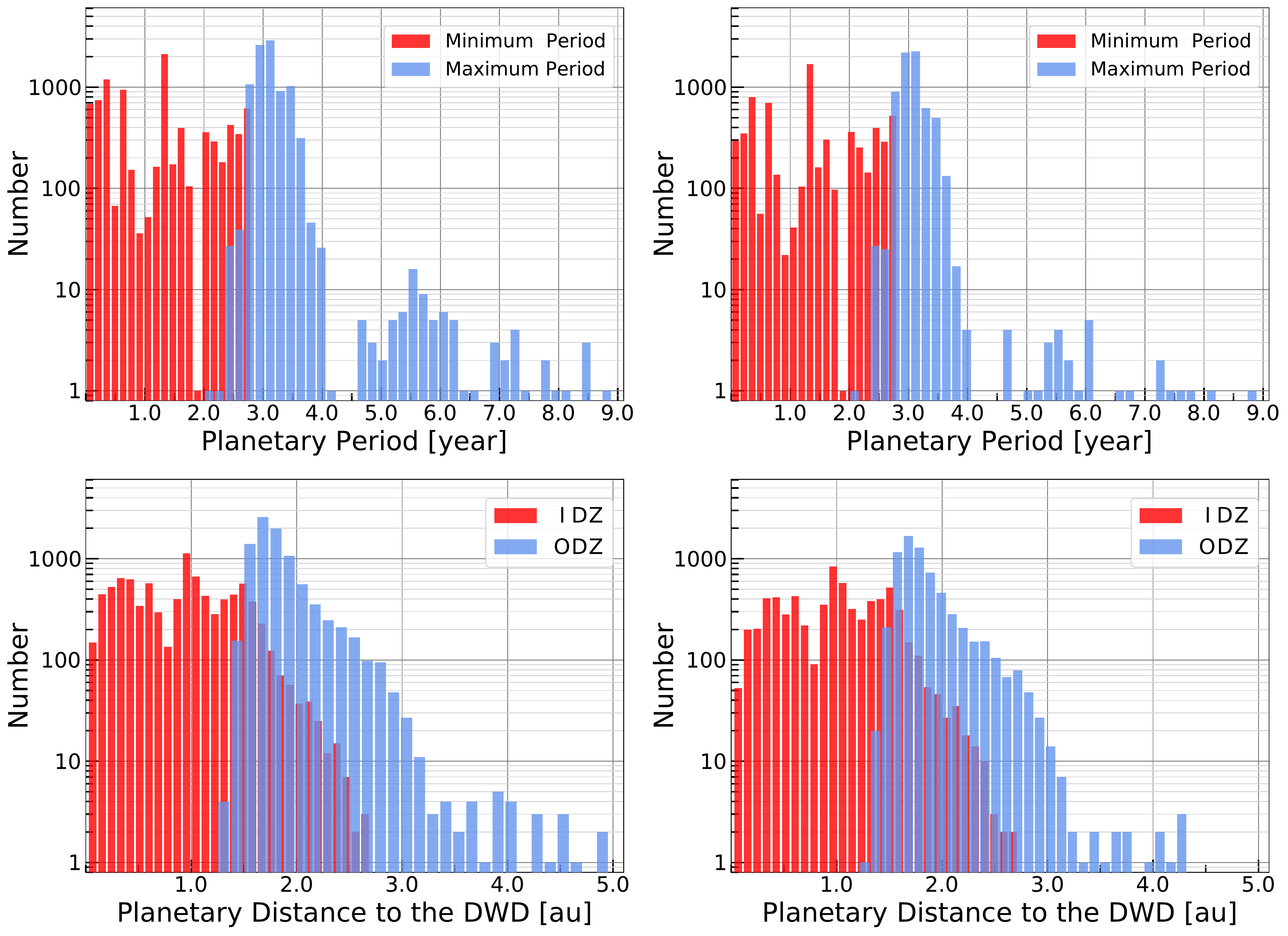} 
    \caption{The distributions of detectable minimum and maximum period (top
    panels), and corresponding IDZ and ODZ (bottom panels). Left panels are for
    Taiji while right panels are  for LISA.}

    \label{ fig:Distribution_2 }
\end{figure*}

To go a step further, we plot the distributions of the detectable minimum
(maximum) period and corresponding IDZ (ODZ) in Fig.~\ref{ fig:Distribution_2 }.
We see that some valleys appear at multiples of one year, which correspond to
the peaks in Fig.~\ref{ fig:Different mass }, caused by the degeneracies between
source and detector parameters (see Sec.~\ref{ sec:4.1 }). From the bottom
panels, we see that Taiji is expected to detect CBPs with the planetary orbital
size smaller than 5\,au, which is about the distance from the Sun to the Jupiter
(5.2\,au), while it is a little smaller (4.4\,au) for LISA. The constraints on
DZs actually reflect the detecting ability of the space-borne GW detector we
choose, because these results are based on the population of DWDs without
injecting any CBP models. It shows that the best range of detection is between
0.1\,au to 3\,au for both LISA and Taiji.

\begin{figure*}
    \centering
    \includegraphics[width=0.8\linewidth]{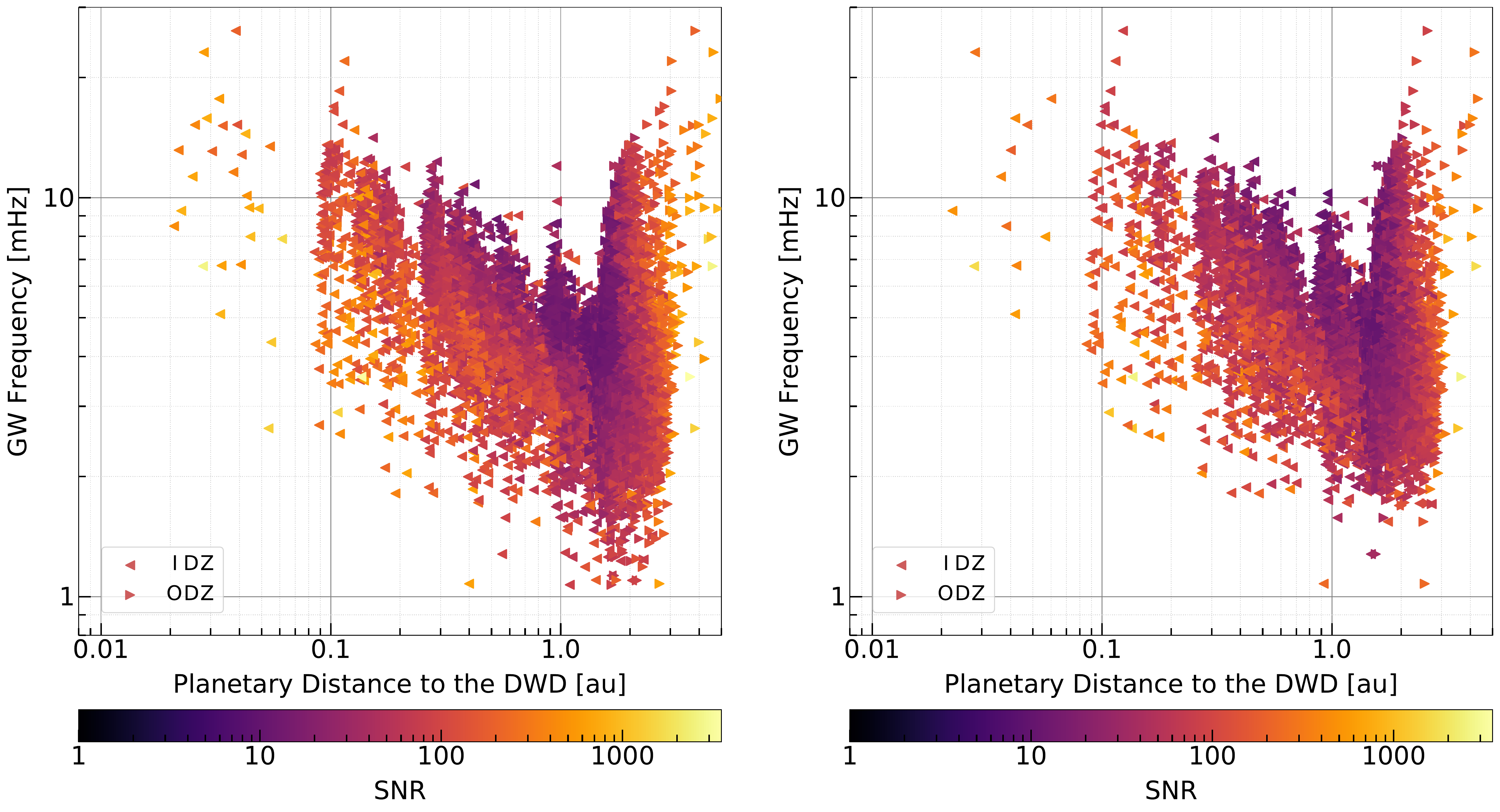}
    \caption{Dependence of IDZ and ODZ on the GW frequency of each promising
    system. The color represents the SNR of each system for Taiji (left) and
    LISA (right). We use the left and right triangles to denote the IDZ and ODZ
    respectively. Notice that for each system, its IDZ and ODZ are located in
    the same horizontal line (i.e. they have the same GW frequency).}

    \label{ fig:Distribution_3 }
\end{figure*}

For each promising system, the dependence of IDZ and ODZ on the GW frequency
are illustrated in Fig.~\ref{ fig:Distribution_3 }. Our results seem to suggest
that systems with higher GW frequencies tend to have wider DZs. An explanation
for this may come from the higher SNRs in the sensitive frequency band, namely
0.1\,mHz to 1\,Hz for LISA and Taiji (see Sec.~\ref{ sec:2.1 } and Fig.~\ref{
fig:SNR }). Note that there are some gaps at distance $\lesssim 1$\,au in
Fig.~\ref{ fig:Distribution_3 }, which is due to our sample intervals in the
parameter estimation process and would not alter the qualitative conclusion
derived here.

\subsubsection{ Detection rates for different CBP models }
\label{ sec:different distributions }

As mentioned in Sec.~\ref{ sec:Exoplanet injection }, we only consider the
coplanar orbits and the occurrence rate is set to 50\% for the promising
detectable systems. Based on the catalog of approximately $2.5 \times 10^4$
($1.6 \times 10^4$) detectable detached DWDs for Taiji (LISA) in total  (see
Sec.~\ref{ sec:4.2.2 }), our model predicts that the total number of the
injected CBP population is 4,507 (3,322) during the nominal mission span. 

{Following \citet{Danielski:2019rvt}, we consider two scenarios in our sub-stellar object (SSO) injection processes:} 
\begin{enumerate}[(A)]
    \item {an optimistic case where $a$ follows a log-uniform distribution in the range of $0.1$\,--\,$200\,\mathrm{au}$, and $M_\mathrm{p}$ is uniformly distributed in the range of $1\,{M}_{\oplus}$\,--\,$0.08\,{M_{\odot}}$, and}
    \item {a pessimistic scenario where $a$ is uniformly distributed in the range of $0.1$\,--\,$200\,\mathrm{au}$, and $M_\mathrm{p}$ is uniformly distributed in the range of $1\,{M}_{\oplus}$\,--\,$0.08\,{M_{\odot}}$.} 
\end{enumerate}

{Note that samples with an injected SSO mass $13\,M_{\mathrm{J}} < M < 0.08\,{M_{\odot}}$ are discarded because we only focus on the CBPs with mass $M_{\mathrm{p}} \leq 13\,M_{\mathrm{J}}$ in this work. More discussions on brown dwarfs with mass $13\,M_{\mathrm{J}} < M < 0.08\,{M_{\odot}}$ can be found in \citet{Danielski:2019rvt}}
    
{We find a total of 40 (16) detected CBPs for (A), 2 (0) for (B), corresponding to 0.16\% (0.10\%) and 0.008\% (0\%) of the total population of detected DWDs over the 4-year mission of Taiji (LISA). From these we conclude that the detection rates in our work are essentially in agreement with the results in \citet{Danielski:2019rvt}, but a little bit more pessimistic as a whole due to the different underlying models and assumptions. Therefore, it is advantageous to improve CBP models in future for more comparisons. Although there seems like no detection in scenario (B) for LISA, Taiji can still has a non-zero result on CBP detections. These data again suggest that Taiji has a better performance on detecting abilities.}

\subsubsection{ Prospects for detections of CBPs in habitable zone around DWDs }
\label{ sec:HZ }

The discovery of thousands of exoplanets in past decades has been promoting the
study of habitability and the search for extraterrestrial life
\citep{2016AsBio..16...89C,2017ARA&A..55..433K, Lingam:2018uva}, which encompass
various research methods within the physical, biological, and environmental
sciences. Among many contemporary habitability metrics, the habitable zone (HZ)
forms a fundamental component to assess the potential habitability of newly
discovered exoplanets. It describes the circumstellar distance where water at
the surface of an exoplanet would be in the liquid phase \citep{Kasting:1993zz},
mainly because all life on the Earth requires liquid water directly or
indirectly.  Given that the Earth is the only known planet with life on it, it
is reasonable to suppose that such a concept also applies to exoplanets beyond
the Earth.

Most research about HZ has focused on MS stars that are similar to the Sun
\citep{Kasting:1993zz, 2007A&A...476.1373S, Lunine:2008iw, 2013AsBio..13..833R}.
But recent studies start to discuss the HZ of WDs \citep{2010BASBr..29...22M,
Agol:2011wx, Fossati:2012kj, 2013AsBio..13..279B}. Although, unlike
hydrogen-burning stars, the WD cooling makes the HZ moves inwards with time, WDs
are still expected to provide a source of energy for planets in HZ for giga-year
(Gyr) durations. As the remnants of MS stars, WDs are as abundant as Sun-like
stars in our Galaxy. Most of them are close in size to our Earth with a
characteristic luminosity of $\sim 10^{-4}\,L_\odot$ \citep{Agol:2011wx}. So the
HZ around WDs is located within $\sim 0.01$\,au where planets must have migrated
inwards after the CE phases \citep{Debes:2002bx, Livio:2005dt,
2011MNRAS.410..899F}. As noted by \citet{Tamanini:2018cqb}, the detection of
such an exoplanet would help to provide crucial information on migration
theories, especially around post-CE binaries.

We assume that the HZ boundary estimations for MS stars also apply to DWD
systems. Thus we can determine the inner/outer edge of the HZ (referred to as
IHZ/OHZ) via equations in \citet{2007A&A...476.1373S}, 
\begin{equation}
\begin{split}
\mathrm{IHZ} &= (\mathrm{IHZ}_\odot-a_{\mathrm{in}} T_{\star}-b_{\mathrm{in}} T_{\star}^{2})\big(\frac{L}{L_{\odot}}\big)^{1/2}\,,\\
\mathrm{OHZ} &= (\mathrm{OHZ}_\odot-a_{\mathrm{out}} T_{\star}-b_{\mathrm{out}} T_{\star}^{2})\big(\frac{L}{L_{\odot}}\big)^{1/2}\,,
\label{ eq:HZ }
\end{split}
\end{equation}
where $\mathrm{IHZ}_\odot$ and $\mathrm{OHZ}_\odot$ are the boundaries in our
Solar system depending on different fractional cloud cover on the day side of an
exoplanet (see Table~\ref{ tab:HZ }). As noted in the Sec.~2 of
\citet{2007A&A...476.1373S}, clouds can increase the planetary albedo and reduce
the greenhouse warming, which thus moves $\mathrm{IHZ}_\odot$ closer to the
star. But for $\mathrm{OHZ}_\odot$ associated with $\mathrm{CO_2}$-ice clouds,
which differ significantly from $\mathrm{H_2O}$-ice particles in the optical
properties, the cooling effect caused by the increase of albedo is weaker than
the warming effect caused by the backscattering of the infrared surface emission
\citep{Lunine:2008iw}. As a result, the theoretical $\mathrm{OHZ}_\odot$ should
be farther for a 100\% cloud cover. Other empirically determined constants in
Eq.~(\ref{ eq:HZ }) are,
\begin{align}
    a_{\mathrm{in}} &= 2.7619\times10^{-5}\,{\rm au \, K^{-1}} \,, \nonumber \\
    b_{\mathrm{in}} &= 3.8095\times10^{-9}\,\mathrm{au \, K^{-2}} \,, \nonumber \\
    a_{\mathrm{out}} &= 1.3786\times10^{-4} \, {\rm au\, K^{-1}} \,, \nonumber \\
    b_{\mathrm{out}} &= 1.4286\times10^{-9}\, \mathrm{au \, K^{-2}} \,.
\end{align}
In Eq.~(\ref{ eq:HZ }), $L$ and $L_{\odot}$ are the primary's and Solar
luminosity, respectively, and $T_{\star}=T_{\mathrm{eff}}-5700$\,K, where the
effective temperature of the primary $T_{\mathrm{eff}}$ is given according to
the Stefan-Boltzmann law via,
\begin{equation}
T_\mathrm{e f f}=\left(\frac{L}{4 \pi \sigma R^{2}}\right)^{1/4}\,,
\end{equation}
with $\sigma$ being the Stefan-Boltzmann constant and $R$ being the primary's
radius.

\begin{table*}[ht]
    \centering
    \caption{Detection numbers and percentages of habitable CBPs around the
    promising systems for LISA and Taiji. We have mentioned in Sec.~\ref{
    sec:4.2.2 } that the total number of promising systems is 9,053 for Taiji,
    and 6,718 for LISA  for a 4-yr mission.}
    \setlength{\tabcolsep}{1.8mm}{\begin{tabular}{c cccc cccc cccc}
    \toprule
                 & \multicolumn{4}{c}{Clouds 0\%}
                 & \multicolumn{4}{c}{Clouds 50\%}    
                 & \multicolumn{4}{c}{Clouds 100\%}
                 \vspace{-0.5em}\\
    Detector    & $\mathrm{IHZ}_\odot$
                 & $\mathrm{OHZ}_\odot$
                 & IHZ    & OHZ        
                 
                 & \hspace{1.5em}$\mathrm{IHZ}_\odot$
                 & $\mathrm{OHZ}_\odot$
                 & IHZ    & OHZ   
                 & \hspace{1.4em}$\mathrm{IHZ}_\odot$
                 & $\mathrm{OHZ}_\odot$
                 & \hspace{-0.2em}IHZ    & \hspace{-0.1em}OHZ  
                 \vspace{-0.5em}\\
                 & \hspace{-0.4em}[au]    & \hspace{-0.4em}[au]    
                 & [au]    & [au]        
                 & \hspace{1.1em}[au]    & \hspace{-0.4em}[au]    
                 & [au]    & [au]
                 & \hspace{1.1em}[au]    & \hspace{-0.4em}[au]    
                 & [au]    & [au]    
                 \vspace{0.3em}\\
    \midrule 
                 & \hspace{-0.4em}0.895    & \hspace{-0.4em}1.67    & 0.013    &  \hspace{0.2em}0.024
                 & \hspace{1em}0.72     & 1.95    & 0.010    &  0.028
                 & \hspace{1.2em}0.485    & \hspace{-0.4em}2.4     & 0.0069   &  0.034
                 \\
    Taiji        & \multicolumn{4}{c}{\hspace{0.2em}$3\quad(0.033\%)$}
                 & \multicolumn{4}{c}{\hspace{0.5em}$5\quad(0.055\%)$}
                 & \multicolumn{4}{c}{\hspace{0.2em}$13\quad(0.14\%)$}
                 \\
    \hspace{0.2em}LISA         
                 & \multicolumn{4}{c}{\hspace{0.2em}$1\quad(0.015\%)$}
                 & \multicolumn{4}{c}{\hspace{0.5em}$1\quad(0.015\%)$}
                 & \multicolumn{4}{c}{\hspace{0.7em}$3\quad(0.45\%)$}
                 \vspace{0.3em}\\ 
    \bottomrule
    \end{tabular}}
    \label{ tab:HZ }
\end{table*}

As noted by \citet{2013AsBio..13..279B}, WDs cool rapidly for about 3\,Gyr, and
then maintain a relatively constant temperature before falling off again at
about 7\,Gyr. As a rough estimation, we neglect the distributions of
cooling time and assume that each WD has the same fixed luminosity value for
which we set it to $1\times10^{-4}$\,$L_{\odot}$, with a  total luminosity 
$2\times10^{-4}$\,$L_{\odot}$. Through the mass-radius relation for WDs in the
Newtonian case, as provided in Fig.~4 of \citet{Ambrosino:2020xoe}, we calculate
$T_{\mathrm{eff}}$ of each promising system. Based on our assumption, the two
WDs have the same mass in the DWD system, therefore their $T_{\mathrm{eff}}$
values are the same as well. We plug $T_{\mathrm{eff}}$ into Eq.~(\ref{ eq:HZ })
to obtain the IHZ and OHZ of each promising system. We verify that ODZ is much
farther than OHZ for each  system due to the low luminosities of WDs, which
means that we only need to compare the limits between OHZ and IDZ. If the IDZ
lies closer to the DWD than the OHZ, we can say that LISA and Taiji are possible
to detect a habitable CBP around this system. 

We list our results in Table~\ref{ tab:HZ } for LISA and Taiji during a 4-year
observation.  Note that for each cloud cover scheme, we take the same values of
IHZ and OHZ as a criterion because the boundary values of HZ are insensitive to
$T_{\mathrm{eff}}$ in our simplified model.  Although the detection numbers do
not seem many, it still shows that such a possibility exists. We also have
verified that the detection number would not increase by more than one order of
magnitude if we set the total luminosity to $2\times10^{-3}\,L_{\odot}$, which
is larger than what they really are when the WD cooling is considered \citep[see
e.g.\ Fig.~1 in][]{2013AsBio..13..279B}.

\section{ Conclusion }
\label{ sec:Conclusion }

In this work, we introduce the two space-borne GW detectors, LISA and Taiji, and
discuss the use of them to detect exoplanets.  For the GW detection method
originally proposed by \citet{Tamanini:2018cqb}, we give some complementary
calculations on the detecting abilities with different values of binary mass and
mass ratio. The conceptual idea using LISA/Taiji to detect exoplanets  is
similar to the RV technique but has unique advantages, compared to the
traditional EM methods.  Before quantitatively analyzing the prospects for
detecting CBPs around DWDs in the whole Galaxy by using LISA and Taiji, we show
that there is a possibility to detect CBPs around four known detached DWDs with
high SNRs using Taiji, while three systems are also promising for LISA. The
minimum detectable masses around these DWDs can be as small as a few of the
Jupiter mass. Moreover, if EM observations can give more constraints on these
systems in the meantime, e.g.~inferring the orbital inclination of the CBP, GWs
may place more restrictions on the mass of the CBP and the existence of such a
population. 

Based on the DWD population from MLDC Round~4, we give quick assessments of CBP
detections in the whole Galaxy during a 4-yr mission time of LISA/Taiji. Our
results show that LISA can detect $\sim 6,000$ promising systems, while the
number rises to $\sim 9,000$ for Taiji. From the distributions of DZs, we show
that the best range of CBP detections is between 0.1\,au to 3\,au around DWDs
for both LISA and Taiji. {Furthermore, we inject two different planet distributions with an occurrence rate of 50\%, following
\citet{Danielski:2019rvt}, to constrain the total detection rates. Our results
are in bold agreement with previous studies, but seem slightly more pessimistic as a whole due to the different models and assumptions we adopted.} By assuming that the HZ boundary estimations for MS stars also apply to
DWDs, we briefly discuss the prospects for detecting habitable CBPs around
detached DWDs in a simplified model. It shows that such a possibility exists
though the detection rates are not large during 4-yr observations.

In addition to the planetary migration theories \citep{2015ExA....40..501T},
there are also some studies about the second-generation formation process which
can be used to explain the existence of nearby exoplanets in these systems
\citep{Zorotovic:2012aa, 2014A&A...562A..19V, 2014A&A...563A..61S}. All our
results can actually help analyze planetary systems after CE phases and provide
a useful input for exoplanetary projects. With a rapid development of GW
astronomy in the past 5 years, we look forward to the synergy with EM
observations and the full investigation of such a GW detection method of
exoplanets in the near future.

\acknowledgments

We thank the anonymous referee for suggestions.
This work was supported by the National Natural Science Foundation of China
(11975027, 11991053, 11721303),  the National SKA Program of China
(2020SKA0120300), the Young Elite Scientists Sponsorship Program by the China
Association for Science and Technology (2018QNRC001), the Max Planck Partner
Group Program funded by the Max Planck Society, and the High-Performance
Computing Platform of Peking University. YK acknowledges the Hui-Chun Chin and
Tsung-Dao Lee Chinese Undergraduate Research Endowment (Chun-Tsung Endowment) at
Peking University.  This research has made use of the NASA Exoplanet Archive,
which is operated by the California Institute of Technology, under contract with
the National Aeronautics and Space Administration under the Exoplanet
Exploration Program.

\vspace{5mm}
\facilities{LISA, Taiji, Exoplanet Archive}

\appendix

All of the selected known detached DWDs are given in Table \ref{ tab:known DWDs
}.  We list the heavier and the lighter masses of the DWD $m_1$ and $m_2$,
the GW frequency $f$, the luminosity distance $d_\mathrm{DWD}$, the ecliptic
coordinates ($\lambda$, $\beta$), and the inclination angle $\iota$. Most of
these parameters above are taken directly from \citet{Huang:2020rjf}, except the
distance to ZTF~J153932.16+502738.8 (with an asterisk). We have corrected it
with the result in \citet{Burdge:2019hgl}.

\begin{table*}[ht]\normalsize
\centering
\caption{Properties of the selected known detached DWDs \citep{Huang:2020rjf}.
Some inclination angles are given with a square bracket due to lack of direct
measurements of them.  These estimated values are assigned based on the
evolutionary stage and the mass ratio of the system. The asterisk marks a
corrected distance from \citet{Burdge:2019hgl}.}
\setlength{\tabcolsep}{4mm}{\begin{tabular}{l r@{.}l r@{.}l cc r@{.}l r@{.}l r@{.}l}
\toprule
Source
          & \multicolumn{2}{c}{\hspace{-0.2em}$m_1$}    
          & \multicolumn{2}{c}{\hspace{-0.2em}$m_2$}
          & $f$  
          & \hspace{0.2em}$d_\mathrm{DWD}$            
          & \multicolumn{2}{c}{\hspace{0.2em}$\lambda$}            
          & \multicolumn{2}{c}{\hspace{0.2em}$\beta$}    
          & \multicolumn{2}{c}{\hspace{0.1em}$\iota$}                         
          \vspace{-0.4em}\\
          
          & \multicolumn{2}{c}{\hspace{-0.2em}[$\mathrm{M_{\odot}}$]}  
          & \multicolumn{2}{c}{[$\mathrm{M_{\odot}}$]}    & [mHz]  
          & \hspace{0.2em}[kpc]       & \multicolumn{2}{c}{\hspace{0.2em}[deg]}                   
          & \multicolumn{2}{c}{\hspace{0.2em}[deg]}   
          & \multicolumn{2}{c}{\hspace{0.1em}[deg]}                                        
          \vspace{0.25em}\\
\midrule

ZTF~J153932.16+502738.8      & 0 & 61    & 0 & 21    & 4.82  
                             & \hspace{0.2em}2.34*  
                             & \hspace{0.2em}205 & 03  
                             & 66 & 16   & 84 & 0    \\
SDSS~J065133.34+284423.4     & 0 & 49    & 0 & 247   & 2.61      
                             & \hspace{0.2em}0.933  
                             & \hspace{0.2em}101 & 34  
                             & 5 & 80    & 86 & 9   \\
SDSS~J093506.92+441107.0     & 0 & 75    & 0 & 312   & 1.68      
                             & \hspace{0.2em}0.645 
                             & \hspace{0.2em}130 & 98  
                             & 28  & 09  & [60 & 0]         \\
SDSS~J232230.20+050942.06    & 0 & 27    & 0 & 24    & 1.66      
                             & \hspace{0.2em}0.779  
                             & \hspace{0.2em}353 & 44  
                             & 8  & 46   & 27 & 0   \\
PTF~J053332.05+020911.6      & 0 & 65    & 0 & 167   & 1.62      
                             & \hspace{0.2em}1.253  
                             & \hspace{0.2em}82 & 91  
                             & $-21$ & 12  & 72 & 8   \\
SDSS~J163030.58+423305.7     & 0 & 76    & 0 & 298   & 0.84      
                             & \hspace{0.2em}1.019  
                             & \hspace{0.2em}231 & 76  
                             & 63  & 05  & [60 & 0]   \\
SDSS~J092345.59+302805.0     & 0 & 76    & 0 & 275   & 0.51      
                             & \hspace{0.2em}0.299 
                             & \hspace{0.2em}133 & 72 
                             & 14  & 43  & [60 & 0]   \\
\bottomrule
\end{tabular}}
\label{ tab:known DWDs }
\end{table*}

\bibliography{refs}{}

\begin{thebibliography}{}
\expandafter\ifx\csname natexlab\endcsname\relax\def\natexlab#1{#1}\fi
\providecommand{\url}[1]{\href{#1}{#1}}
\providecommand{\dodoi}[1]{doi:~\href{http://doi.org/#1}{\nolinkurl{#1}}}
\providecommand{\doeprint}[1]{\href{http://ascl.net/#1}{\nolinkurl{http://ascl.net/#1}}}
\providecommand{\doarXiv}[1]{\href{https://arxiv.org/abs/#1}{\nolinkurl{https://arxiv.org/abs/#1}}}

\bibitem[{Abbott {et~al.}(2016)}]{Abbott:2016blz}
Abbott, B.~P., {et~al.} 2016, PhRvL, 116, 061102,
  \dodoi{10.1103/PhysRevLett.116.061102}

\bibitem[{Abbott {et~al.}(2017)}]{TheLIGOScientific:2017qsa}
---. 2017, PhRvL, 119, 161101, \dodoi{10.1103/PhysRevLett.119.161101}

\bibitem[{Agol(2011)}]{Agol:2011wx}
Agol, E. 2011, ApJL, 731, L31, \dodoi{10.1088/2041-8205/731/2/L31}

\bibitem[{Althaus {et~al.}(2010)Althaus, Corsico, Isern, \&
  a~Berro}]{Althaus:2010pi}
Althaus, L.~G., Corsico, A.~H., Isern, J., \& a~Berro, E.~G. 2010, A\&ARv, 18,
  471, \dodoi{10.1007/s00159-010-0033-1}

\bibitem[{Amaro-Seoane {et~al.}(2017)}]{Audley:2017drz}
Amaro-Seoane, P., {et~al.} 2017.
\newblock \doarXiv{1702.00786}

\bibitem[{Ambrosino(2020)}]{Ambrosino:2020xoe}
Ambrosino, F. 2020.
\newblock \doarXiv{2012.01242}

\bibitem[{Babak {et~al.}(2008)}]{Babak:2008aa}
Babak, S., {et~al.} 2008, CQGra, 25, 184026,
  \dodoi{10.1088/0264-9381/25/18/184026}

\bibitem[{Babak {et~al.}(2010)}]{Babak:2009cj}
---. 2010, CQGra, 27, 084009, \dodoi{10.1088/0264-9381/27/8/084009}

\bibitem[{{Barnes} \& {Heller}(2013)}]{2013AsBio..13..279B}
{Barnes}, R., \& {Heller}, R. 2013, AsBio, 13, 279,
  \dodoi{10.1089/ast.2012.0867}

\bibitem[{Berti {et~al.}(2005)Berti, Buonanno, \& Will}]{Berti:2004bd}
Berti, E., Buonanno, A., \& Will, C.~M. 2005, PhRvD, 71, 084025,
  \dodoi{10.1103/PhysRevD.71.084025}

\bibitem[{Beuermann {et~al.}(2010)}]{Beuermann:2010bt}
Beuermann, K., {et~al.} 2010, A\&A, 521, L60,
  \dodoi{10.1051/0004-6361/201015472}

\bibitem[{Beuermann {et~al.}(2011)}]{Beuermann:2010ny}
---. 2011, A\&A, 526, A53, \dodoi{10.1051/0004-6361/201015942}

\bibitem[{Breivik {et~al.}(2020)}]{Breivik:2019lmt}
Breivik, K., {et~al.} 2020, ApJ, 898, 71, \dodoi{10.3847/1538-4357/ab9d85}

\bibitem[{{Brown} {et~al.}(2017){Brown}, {Veras}, \&
  {G{\"a}nsicke}}]{2017MNRAS.468.1575B}
{Brown}, J.~C., {Veras}, D., \& {G{\"a}nsicke}, B.~T. 2017, \mnras, 468, 1575,
  \dodoi{10.1093/mnras/stx428}

\bibitem[{{Brown} {et~al.}(2020){Brown}, {Kilic}, {Kosakowski}, {Andrews},
  {Heinke}, {Ag{\"u}eros}, {Camilo}, {Gianninas}, {Hermes}, \&
  {Kenyon}}]{2020ApJ...889...49B}
{Brown}, W.~R., {Kilic}, M., {Kosakowski}, A., {et~al.} 2020, \apj, 889, 49,
  \dodoi{10.3847/1538-4357/ab63cd}

\bibitem[{Burdge {et~al.}(2019)}]{Burdge:2019hgl}
Burdge, K.~B., {et~al.} 2019, Natur, 571, 528,
  \dodoi{10.1038/s41586-019-1403-0}

\bibitem[{{Cockell} {et~al.}(2016){Cockell}, {Bush}, {Bryce}, {Direito},
  {Fox-Powell}, {Harrison}, {Lammer}, {Landenmark}, {Martin-Torres},
  {Nicholson}, {Noack}, {O'Malley-James}, {Payler}, {Rushby}, {Samuels},
  {Schwendner}, {Wadsworth}, \& {Zorzano}}]{2016AsBio..16...89C}
{Cockell}, C.~S., {Bush}, T., {Bryce}, C., {et~al.} 2016, AsBio, 16, 89,
  \dodoi{10.1089/ast.2015.1295}

\bibitem[{Cornish \& Larson(2003)}]{Cornish:2003vj}
Cornish, N.~J., \& Larson, S.~L. 2003, PhRvD, 67, 103001,
  \dodoi{10.1103/PhysRevD.67.103001}

\bibitem[{{Cutler}(1998)}]{1998PhRvD..57.7089C}
{Cutler}, C. 1998, \prd, 57, 7089, \dodoi{10.1103/PhysRevD.57.7089}

\bibitem[{{Cutler} \& {Thorne}(2002)}]{2002grg..conf...72C}
{Cutler}, C., \& {Thorne}, K.~S. 2002, in General Relativity and Gravitation,
  ed. N.~T. {Bishop} \& S.~D. {Maharaj}, 72--111,
  \dodoi{10.1142/9789812776556\_0004}

\bibitem[{Danielski {et~al.}(2019)Danielski, Korol, Tamanini, \&
  Rossi}]{Danielski:2019rvt}
Danielski, C., Korol, V., Tamanini, N., \& Rossi, E.~M. 2019, A\&A, 632, A113,
  \dodoi{10.1051/0004-6361/201936729}

\bibitem[{Danielski \& Tamanini(2020)}]{Danielski:2020hxb}
Danielski, C., \& Tamanini, N. 2020, IJMPD, 29, 2043007,
  \dodoi{10.1142/S0218271820430075}

\bibitem[{Debes \& Sigurdsson(2002)}]{Debes:2002bx}
Debes, J.~H., \& Sigurdsson, S. 2002, ApJ, 572, 556, \dodoi{10.1086/340291}

\bibitem[{Duch\^ene \& Kraus(2013)}]{Duchene:2013cba}
Duch\^ene, G., \& Kraus, A. 2013, ARA\&A, 51, 269,
  \dodoi{10.1146/annurev-astro-081710-102602}

\bibitem[{{Duncan} \& {Lissauer}(1998)}]{1998Icar..134..303D}
{Duncan}, M.~J., \& {Lissauer}, J.~J. 1998, Icar, 134, 303,
  \dodoi{10.1006/icar.1998.5962}

\bibitem[{{Dvorak}(1986)}]{1986A&A...167..379D}
{Dvorak}, R. 1986, \aap, 167, 379

\bibitem[{{Eberle} {et~al.}(2008){Eberle}, {Cuntz}, \&
  {Musielak}}]{2008A&A...489.1329E}
{Eberle}, J., {Cuntz}, M., \& {Musielak}, Z.~E. 2008, \aap, 489, 1329,
  \dodoi{10.1051/0004-6361:200809758}

\bibitem[{{Faedi} {et~al.}(2011){Faedi}, {West}, {Burleigh}, {Goad}, \&
  {Hebb}}]{2011MNRAS.410..899F}
{Faedi}, F., {West}, R.~G., {Burleigh}, M.~R., {Goad}, M.~R., \& {Hebb}, L.
  2011, \mnras, 410, 899, \dodoi{10.1111/j.1365-2966.2010.17488.x}

\bibitem[{{Farihi}(2016)}]{2016NewAR..71....9F}
{Farihi}, J. 2016, \nar, 71, 9, \dodoi{10.1016/j.newar.2016.03.001}

\bibitem[{{Farihi} {et~al.}(2018){Farihi}, {van Lieshout}, {Cauley}, {Dennihy},
  {Su}, {Kenyon}, {Wilson}, {Toloza}, {G{\"a}nsicke}, {von Hippel}, {Redfield},
  {Debes}, {Xu}, {Rogers}, {Bonsor}, {Swan}, {Pala}, \&
  {Reach}}]{2018MNRAS.481.2601F}
{Farihi}, J., {van Lieshout}, R., {Cauley}, P.~W., {et~al.} 2018, \mnras, 481,
  2601, \dodoi{10.1093/mnras/sty2331}

\bibitem[{Fossati {et~al.}(2012)Fossati, Bagnulo, Haswell, Patel, Busuttil,
  Kowalski, Shulyak, \& Sterzik}]{Fossati:2012kj}
Fossati, L., Bagnulo, S., Haswell, C.~A., {et~al.} 2012, ApJL, 757, L15,
  \dodoi{10.1088/2041-8205/757/1/L15}

\bibitem[{Foucart \& Lai(2013)}]{Foucart:2012xe}
Foucart, F., \& Lai, D. 2013, ApJ, 764, 106,
  \dodoi{10.1088/0004-637X/764/1/106}

\bibitem[{{Holman} \& {Wiegert}(1999)}]{1999AJ....117..621H}
{Holman}, M.~J., \& {Wiegert}, P.~A. 1999, \aj, 117, 621,
  \dodoi{10.1086/300695}

\bibitem[{{Hong} \& {van Putten}(2021)}]{2021NewA...8401516H}
{Hong}, C., \& {van Putten}, M. H.~P.~M. 2021, \na, 84, 101516,
  \dodoi{10.1016/j.newast.2020.101516}

\bibitem[{Huang {et~al.}(2020)Huang, Hu, Korol, Li, Liang, Lu, Wang, Yu, \&
  Mei}]{Huang:2020rjf}
Huang, S.-J., Hu, Y.-M., Korol, V., {et~al.} 2020, PhRvD, 102, 063021,
  \dodoi{10.1103/PhysRevD.102.063021}

\bibitem[{Jura {et~al.}(2009)Jura, Farihi, \& Zuckerman}]{Jura:2008qm}
Jura, M., Farihi, J., \& Zuckerman, B. 2009, AJ, 137, 3191,
  \dodoi{10.1088/0004-6256/137/2/3191}

\bibitem[{{Kaltenegger}(2017)}]{2017ARA&A..55..433K}
{Kaltenegger}, L. 2017, \araa, 55, 433,
  \dodoi{10.1146/annurev-astro-082214-122238}

\bibitem[{Kasting {et~al.}(1993)Kasting, Whitmire, \&
  Reynolds}]{Kasting:1993zz}
Kasting, J.~F., Whitmire, D.~P., \& Reynolds, R.~T. 1993, Icar, 101, 108,
  \dodoi{10.1006/icar.1993.1010}

\bibitem[{Kennedy {et~al.}(2012)Kennedy, Wyatt, Sibthorpe, Phillips, Matthews,
  \& Greaves}]{Kennedy:2012gc}
Kennedy, G.~M., Wyatt, M.~C., Sibthorpe, B., {et~al.} 2012, MNRAS, 426, 2115,
  \dodoi{10.1111/j.1365-2966.2012.21865.x}

\bibitem[{{Klein} {et~al.}(2016){Klein}, {Barausse}, {Sesana}, {Petiteau},
  {Berti}, {Babak}, {Gair}, {Aoudia}, {Hinder}, {Ohme}, \&
  {Wardell}}]{2016PhRvD..93b4003K}
{Klein}, A., {Barausse}, E., {Sesana}, A., {et~al.} 2016, \prd, 93, 024003,
  \dodoi{10.1103/PhysRevD.93.024003}

\bibitem[{Koester {et~al.}(2014)Koester, Gänsicke, \&
  Farihi}]{2014A&A...566A..34K}
Koester, D., Gänsicke, B.~T., \& Farihi, J. 2014, A\&A, 566, A34,
  \dodoi{10.1051/0004-6361/201423691}

\bibitem[{Korol {et~al.}(2018)Korol, Koop, \& Rossi}]{Korol:2018ulo}
Korol, V., Koop, O., \& Rossi, E.~M. 2018, ApJL, 866, L20,
  \dodoi{10.3847/2041-8213/aae587}

\bibitem[{Korol {et~al.}(2017)Korol, Rossi, Groot, Nelemans, Toonen, \&
  Brown}]{Korol:2017qcx}
Korol, V., Rossi, E.~M., Groot, P.~J., {et~al.} 2017, MNRAS, 470, 1894,
  \dodoi{10.1093/mnras/stx1285}

\bibitem[{Korol {et~al.}(2020)}]{Korol:2020lpq}
Korol, V., {et~al.} 2020, A\&A, 638, A153, \dodoi{10.1051/0004-6361/202037764}

\bibitem[{Lamberts {et~al.}(2019)Lamberts, Blunt, Littenberg, Garrison-Kimmel,
  Kupfer, \& Sanderson}]{Lamberts:2019nyk}
Lamberts, A., Blunt, S., Littenberg, T.~B., {et~al.} 2019, MNRAS, 490, 5888,
  \dodoi{10.1093/mnras/stz2834}

\bibitem[{Lamberts {et~al.}(2018)Lamberts, Garrison-Kimmel, Hopkins, Quataert,
  Bullock, Faucher-Gigu\`ere, Wetzel, Keres, Drango, \&
  Sanderson}]{Lamberts:2018cge}
Lamberts, A., Garrison-Kimmel, S., Hopkins, P., {et~al.} 2018, MNRAS, 480,
  2704, \dodoi{10.1093/mnras/sty2035}

\bibitem[{Lingam \& Loeb(2018)}]{Lingam:2018uva}
Lingam, M., \& Loeb, A. 2018, JCAP, 05, 020,
  \dodoi{10.1088/1475-7516/2018/05/020}

\bibitem[{Livio {et~al.}(2005)Livio, Pringle, \& Wood}]{Livio:2005dt}
Livio, M., Pringle, J.~E., \& Wood, K. 2005, ApJL, 632, L37,
  \dodoi{10.1086/497577}

\bibitem[{{Livio} \& {Soker}(1984)}]{1984MNRAS.208..763L}
{Livio}, M., \& {Soker}, N. 1984, \mnras, 208, 763,
  \dodoi{10.1093/mnras/208.4.763}

\bibitem[{Lunine {et~al.}(2008)}]{Lunine:2008iw}
Lunine, J.~I., {et~al.} 2008.
\newblock \doarXiv{0808.2754}

\bibitem[{{Luo} {et~al.}(2020){Luo}, {Guo}, {Jin}, {Wu}, \&
  {Hu}}]{2020ResPh..1602918L}
{Luo}, Z., {Guo}, Z., {Jin}, G., {Wu}, Y., \& {Hu}, W. 2020, ResPh, 16, 102918,
  \dodoi{10.1016/j.rinp.2019.102918}

\bibitem[{{Monteiro}(2010)}]{2010BASBr..29...22M}
{Monteiro}, H. 2010, BASBr, 29, 22

\bibitem[{{Mustill} {et~al.}(2018){Mustill}, {Villaver}, {Veras},
  {G{\"a}nsicke}, \& {Bonsor}}]{2018MNRAS.476.3939M}
{Mustill}, A.~J., {Villaver}, E., {Veras}, D., {G{\"a}nsicke}, B.~T., \&
  {Bonsor}, A. 2018, \mnras, 476, 3939, \dodoi{10.1093/mnras/sty446}

\bibitem[{Nelemans \& Tauris(1998)}]{Nelemans:1998axa}
Nelemans, G., \& Tauris, T.~M. 1998, A\&A, 335, L85.
\newblock \doarXiv{astro-ph/9806011}

\bibitem[{Nelemans {et~al.}(2001)Nelemans, Yungelson, \&
  Portegies~Zwart}]{Nelemans:2001hp}
Nelemans, G., Yungelson, L.~R., \& Portegies~Zwart, S.~F. 2001, A\&A, 375, 890,
  \dodoi{10.1051/0004-6361:20010683}

\bibitem[{Raghavan {et~al.}(2010)Raghavan, McAlister, Henry, Latham, Marcy,
  Mason, Gies, White, \& Brummelaar}]{Raghavan:2010hq}
Raghavan, D., McAlister, H.~A., Henry, T.~J., {et~al.} 2010, ApJS, 190, 1,
  \dodoi{10.1088/0067-0049/190/1/1}

\bibitem[{{Ramsay} {et~al.}(2018){Ramsay}, {Green}, {Marsh}, {Kupfer},
  {Breedt}, {Korol}, {Groot}, {Knigge}, {Nelemans}, {Steeghs}, {Woudt}, \&
  {Aungwerojwit}}]{2018A&A...620A.141R}
{Ramsay}, G., {Green}, M.~J., {Marsh}, T.~R., {et~al.} 2018, \aap, 620, A141,
  \dodoi{10.1051/0004-6361/201834261}

\bibitem[{Robson {et~al.}(2019)Robson, Cornish, \& Liu}]{Robson_2019}
Robson, T., Cornish, N.~J., \& Liu, C. 2019, CQGra, 36, 105011,
  \dodoi{10.1088/1361-6382/ab1101}

\bibitem[{Roebber {et~al.}(2020)}]{Roebber:2020hso}
Roebber, E., {et~al.} 2020, ApJL, 894, L15, \dodoi{10.3847/2041-8213/ab8ac9}

\bibitem[{Ruan {et~al.}(2020)Ruan, Liu, Guo, Wu, \& Cai}]{Ruan:2020smc}
Ruan, W.-H., Liu, C., Guo, Z.-K., Wu, Y.-L., \& Cai, R.-G. 2020, NatAs, 4, 108,
  \dodoi{10.1038/s41550-019-1008-4}

\bibitem[{{Rushby} {et~al.}(2013){Rushby}, {Claire}, {Osborn}, \&
  {Watson}}]{2013AsBio..13..833R}
{Rushby}, A.~J., {Claire}, M.~W., {Osborn}, H., \& {Watson}, A.~J. 2013, AsBio,
  13, 833, \dodoi{10.1089/ast.2012.0938}

\bibitem[{{Schleicher} \& {Dreizler}(2014)}]{2014A&A...563A..61S}
{Schleicher}, D. R.~G., \& {Dreizler}, S. 2014, \aap, 563, A61,
  \dodoi{10.1051/0004-6361/201322860}

\bibitem[{{Selsis} {et~al.}(2007){Selsis}, {Kasting}, {Levrard}, {Paillet},
  {Ribas}, \& {Delfosse}}]{2007A&A...476.1373S}
{Selsis}, F., {Kasting}, J.~F., {Levrard}, B., {et~al.} 2007, \aap, 476, 1373,
  \dodoi{10.1051/0004-6361:20078091}

\bibitem[{Seto(2008)}]{Seto:2008di}
Seto, N. 2008, Astrophys. J. Lett., 677, L55, \dodoi{10.1086/587785}

\bibitem[{{Shi} {et~al.}(2019){Shi}, {Bao}, {Wang}, {Zhang}, {Hu}, {Sesana},
  {Barausse}, {Mei}, \& {Luo}}]{2019PhRvD.100d4036S}
{Shi}, C., {Bao}, J., {Wang}, H.-T., {et~al.} 2019, \prd, 100, 044036,
  \dodoi{10.1103/PhysRevD.100.044036}

\bibitem[{{Sigurdsson}(1993)}]{1993ApJ...415L..43S}
{Sigurdsson}, S. 1993, \apjl, 415, L43, \dodoi{10.1086/187028}

\bibitem[{{Smallwood} {et~al.}(2018){Smallwood}, {Martin}, {Livio}, \&
  {Lubow}}]{2018MNRAS.480...57S}
{Smallwood}, J.~L., {Martin}, R.~G., {Livio}, M., \& {Lubow}, S.~H. 2018,
  \mnras, 480, 57, \dodoi{10.1093/mnras/sty1819}

\bibitem[{Takahashi \& Seto(2002)}]{Takahashi:2002ky}
Takahashi, R., \& Seto, N. 2002, ApJ, 575, 1030, \dodoi{10.1086/341483}

\bibitem[{Tamanini \& Danielski(2019)}]{Tamanini:2018cqb}
Tamanini, N., \& Danielski, C. 2019, NatAs, 3, 858,
  \dodoi{10.1038/s41550-019-0807-y}

\bibitem[{{Thorsett} {et~al.}(1993){Thorsett}, {Arzoumanian}, \&
  {Taylor}}]{1993ApJ...412L..33T}
{Thorsett}, S.~E., {Arzoumanian}, Z., \& {Taylor}, J.~H. 1993, \apjl, 412, L33,
  \dodoi{10.1086/186933}

\bibitem[{{Turrini} {et~al.}(2015){Turrini}, {Nelson}, \&
  {Barbieri}}]{2015ExA....40..501T}
{Turrini}, D., {Nelson}, R.~P., \& {Barbieri}, M. 2015, ExA, 40, 501,
  \dodoi{10.1007/s10686-014-9401-6}

\bibitem[{{Veras}(2016)}]{2016RSOS....350571V}
{Veras}, D. 2016, RSOS, 3, 150571, \dodoi{10.1098/rsos.150571}

\bibitem[{Veras \& Tout(2012)}]{Veras:2012yi}
Veras, D., \& Tout, C.~A. 2012, MNRAS, 422, 1648,
  \dodoi{10.1111/j.1365-2966.2012.20741.x}

\bibitem[{Veras {et~al.}(2011)Veras, Wyatt, Mustill, Bonsor, \&
  Eldridge}]{Veras:2011di}
Veras, D., Wyatt, M.~C., Mustill, A.~J., Bonsor, A., \& Eldridge, J.~J. 2011,
  MNRAS, 417, 2104, \dodoi{10.1111/j.1365-2966.2011.19393.x}

\bibitem[{{V{\"o}lschow} {et~al.}(2014){V{\"o}lschow}, {Banerjee}, \&
  {Hessman}}]{2014A&A...562A..19V}
{V{\"o}lschow}, M., {Banerjee}, R., \& {Hessman}, F.~V. 2014, \aap, 562, A19,
  \dodoi{10.1051/0004-6361/201322111}

\bibitem[{Wang \& Han(2021)}]{Wang:2021mou}
Wang, G., \& Han, W.-B. 2021, PhRvD, 103, 064021,
  \dodoi{10.1103/PhysRevD.103.064021}

\bibitem[{Wong {et~al.}(2019)Wong, Berti, Gabella, \&
  Holley-Bockelmann}]{Wong:2018amf}
Wong, K. W.~K., Berti, E., Gabella, W.~E., \& Holley-Bockelmann, K. 2019,
  MNRAS, 483, L33, \dodoi{10.1093/mnrasl/sly208}

\bibitem[{Yu \& Jeffery(2010)}]{Yu:2010fq}
Yu, S., \& Jeffery, C.~S. 2010, A\&A, 521, A85,
  \dodoi{10.1051/0004-6361/201014827}

\bibitem[{Zorotovic \& Schreiber(2013)}]{Zorotovic:2012aa}
Zorotovic, M., \& Schreiber, M.~R. 2013, A\&A, 549, A95,
  \dodoi{10.1051/0004-6361/201220321}

\end{thebibliography}
\bibliographystyle{aasjournal}

\end{document}